\documentclass[aps,prx,superscriptaddress,twocolumn]{revtex4-2}

\usepackage{graphicx}
\usepackage[dvipsnames]{xcolor}
\usepackage{bm}
\usepackage{physics}
\usepackage{amsthm}
\usepackage{mathtools}
\usepackage{hyperref}
\usepackage{bbm}
\usepackage{float}
\usepackage{subcaption}
\usepackage{amssymb}
\usepackage[capitalise]{cleveref}
\definecolor{customOrange}{RGB}{253,164,64}
\definecolor{customBlue}{RGB}{76,89,154}

\usepackage{tikz}

\theoremstyle{definition}

\newtheorem{rst}{Result}

\makeatletter

\newcommand{\Rmnum}[1]{\expandafter\@slowromancap\romannumeral #1@}
\makeatother

\DeclareMathOperator\arctanh{arctanh}

\begin{document}

\title{The Petz recovery map for optical losses}

\author{Jinyan Chen}
\affiliation{Centre for Quantum Technologies, National University of Singapore, 3 Science Drive 2, Singapore 117543}

\author{Minjeong Song}
\affiliation{Centre for Quantum Technologies, National University of Singapore, 3 Science Drive 2, Singapore 117543}

\author{Jared Jia Xuan Chan}
\affiliation{Department of Physics, National University of Singapore, 2 Science Drive 3, Singapore 117542}

\author{Valerio Scarani}
\affiliation{Centre for Quantum Technologies, National University of Singapore, 3 Science Drive 2, Singapore 117543}
\affiliation{Department of Physics, National University of Singapore, 2 Science Drive 3, Singapore 117542}

\date{\today}

\begin{abstract}
Optical systems are a main platform for quantum information processing. A main challenge is information loss due to scattering in unmonitored modes. These losses are modeled as state-independent beam-splitter interactions, with a thermal state (for all practical purposes, the vacuum) in the second input port. The perfect correction of these Gaussian lossy channels with Gaussian operations alone is known to be impossible. In this work, we investigate the Petz recovery map as an approximate recovery. For single mode losses and Gaussian reference states, the Petz map is found to use either a beam-splitter or a state-independent amplifier, depending on the parameters. Then we study the recovery performance on several examples, showing that it is near-optimal among the considered class of protocols. We also obtain more specific comparisons: Petz is always better than just re-preparing the reference state; but it is worse than doing nothing if the reference state is far from the true state. Finally, we extend our study to losses on two modes, and compare the global Petz map to the local implementation on each mode separately.

\end{abstract}

\maketitle

\section{\label{sec:level1}Introduction}

Optical degrees of freedom have been a system of choice to encode quantum information from the start. Most of quantum communication is implemented with light; quantum advantage can be demonstrated with boson sampling \cite{bosonsampling}; schemes for optical quantum computing have been proposed, first with postselection \cite{LOQC} and later with multiplexing \cite{fusion}, and are still very actively pursued; finally, optics is of course a major tool for metrolgy \cite{metrology}. In quantum optics, two vector spaces are used \cite{fabretreps}: the classical one of the modes (superposition in the electromagnetic field from the linearity of Maxwell's equations) and the quantum Fock space (which attaches an infinite-dimensional Hilbert space to each mode, in any orthogonal decomposition). Information can be encoded in either. For instance, polarisation, time-bin or dual-rail schemes encode discrete information in several modes; continuous-variable schemes encode information in different quantum states of even a single mode.

When it comes to \textit{information loss during propagation}, by and large the dominant mechanism in optical systems is the loss of electromagnetic amplitude, simply known as ``losses''. Indeed, propagation usually takes place in linear media, where cross-talk among modes is negligible (for instance, qubits encoded in polarisation modes do not decohere), but scattering into unmonitored modes matters. In an input-output description, losses in mode $a$ can be effectively described as a single \textit{state-independent linear beam-splitter} that scatters some amplitude into an unmonitored mode \cite{scully1997quantum}. Importantly, the way such losses affect the encoded information differs with the encoding. If the information was encoded in modes, the information scattered into unmonitored modes is just lost. By contrast, if the information was encoded in the state of a mode, the state itself is modified by the losses -- in other words, \textit{the information itself is degraded}. This is the situation that we consider in this paper.

The problem has been recognized early on, and error correction schemes have been proposed. It is notably known that perfect error correction of such a Gaussian channel is impossible using only Gaussian operations \cite{niset2009no}. In this paper, we consider \textit{approximate correction of losses using the Petz recovery map}~\cite{petz1986sufficient,petz1988sufficiency}. The Petz map is known to guarantee near-optimal recovery according to several criteria  \cite{nearoptimal1,nearpotimal2,nearoptimal3,nearoptimal4}. It does appear naturally in error correction after the encoding stage \cite{Petzerrorcorrection}, but here we focus on its implementation without that stage. In spite of its widespread theoretical use in quantum data processing \cite{dataprocessing}, fluctuation theorems \cite{kwon-kim,fluctuationtheorem} and other thermodynamical entropies \cite{entropy}, interest in its experimental implementation is growing only now \cite{kwon-zhang,wenhan,leaNMR}, and the case of Gaussian channels was also studied only recently \cite{GaussianPetz2}. 

In this work, we will focus on a specific lossy Gaussian channel, and characterize its Petz recovery map when reference states are Gaussian states. The paper is structured as follows: In \cref{Preliminaries}, we introduce the necessary background on Gaussian channels, Gaussian states, and the Gaussian Petz recovery map. In \cref{model}, we present the lossy channel under consideration and its corresponding Petz recovery map. In \cref{fidelity}, we analyze the recovery performance of Petz recovery map comparing to two other trivial recovery protocols that either keep the noisy state or replace it with the reference state. We show that it is near-optimal within a class of recovery maps. In \cref{multi}, we extend to two mode cases, where we show that using global Petz recovery map is better at recovering correlations. 

\section{\label{Preliminaries}Preliminaries}

\subsection{Gaussian states and Gaussian channels}

We summarize here the notions of Gaussians states and channels needed in this work \cite{Gaussianbook}. Denote the quadrature operators for a $n$-mode field by $\mathbf{r} = (q_1,p_1,q_2,p_2,...,q_n,p_n)^T$, with the canonical commutation relations $\comm{r_j}{r_k} = i \hbar\Omega_{jk}$, where $\Omega \equiv \bigoplus_{j=1}^{n} \left(\begin{smallmatrix}
    0 & 1\\
    -1 & 0
\end{smallmatrix}\right)$ is the symplectic form and $T$ represents the transpose. Throughout the paper, we will adopt the natural units $\hbar:=1$. A Gaussian state is a state, whose Wigner function is a Gaussian function in phase space: 
\begin{equation}
    W(\mathbf{r}) = \frac{1}{(2 \pi)^n \sqrt{\det V}}e^{(\mathbf{r}-\overline{\mathbf{r}})^T V^{-1}(\mathbf{r}-\overline{\mathbf{r}})}, 
\end{equation}
where $\overline{\mathbf{r}} = \langle \mathbf{r}\rangle $ is the mean displacement vector and $V$ is the covariance matrix with entries $V_{ij} = \langle r_ir_j+r_jr_i\rangle-2\langle r_i\rangle\langle r_j\rangle$. Covariance matrices are symmetric by construction, i.e., $V=V^T$; the canonical commutation relations imply the Robertson-Schr{\"o}dinger uncertainty relation $\det V \ge 1$, with equality for pure states \cite{robertson1929uncertainty}. We will often use $(\overline{\mathbf{r}}_{\rho},V_\rho)$ to represent the Gaussian state $\rho$.

A Gaussian channel is a completely positive, trace-preserving map that preserves Gaussianity of states. As such, its action on $(\overline{\mathbf{r}},V)$ is a symplectic transformation
\begin{equation}
    \begin{split}
        \overline{\mathbf{r}} &\to X \overline{\mathbf{r}} +\mathbf{d}, \\
        V & \to X V X^T+Y,
    \end{split}
\end{equation}
where $X$ and $Y=Y^T$ are real matrices that represent the transformation matrix and the noise matrix, respectively, and $\mathbf{d}$ is the displacement vector. The condition of complete positivity (CP) is captured by the relation~\cite{Gaussianinformation}
\begin{align}\label{eq:CP}
 Y + i\Omega -iX\Omega X^T \ge 0
\end{align} between $X$ and $Y$. We will often use $(X_\mathcal{N},Y_\mathcal{N},\mathbf{d}_\mathcal{N})$ to represent the Gaussian channel $\mathcal{N}$. In the case of Gaussian unitaries, $X$ is symplectic ($X\Omega X^T = \Omega$) and $Y=0$. 

\subsection{Petz recovery map}

For a channel $\mathcal{N}$ and a reference state $\sigma$, the Petz recovery map is defined by
\begin{align}
\mathcal{P}_{\mathcal{N},\sigma}(\bullet)&=\sqrt{\sigma}\,\mathcal{N}^\dagger\left(\frac{1}{\sqrt{\mathcal{N}(\sigma)}}\bullet\frac{1}{\sqrt{\mathcal{N}(\sigma)}}\right) \sqrt{\sigma}\,.   
\end{align} The inverse in this expression should be taken as pseudo-inverse, defined only on the support of $\mathcal{N}(\sigma)$. We shall notably avoid the case when $\mathcal{N}(\sigma)$ is pure, i.e.~we shall only consider cases where $\det V_{\mathcal{N}(\sigma)}>1$ \cite{serafini2023qcv}. The role of the reference state $\sigma$ is that of a prior, and indeed the Petz map defines a quantum analog of Bayesian update \cite{Parzygnat2023axiomsretrodiction,minchange}. One of the properties of the map is that the prior is perfectly recovered \cite{petz1986sufficient,petz1988sufficiency}: $\mathcal{P}_{\mathcal{N},\sigma}\circ\mathcal{N}(\sigma)=\sigma$. If the input was a different state $\rho$, its approximate recovery is determined by the decrease of relative entropy between it and $\sigma$, with perfect recovery for the states for which the relative entropy remains the same \cite{hayden2004qmarkov,recoverability}. As last generality, we mention two extreme examples of channels: the Petz recovery map of the identity channel is the identity channel; the one of an erasure channel $\mathcal{N}(\bullet)=\tau\,\textrm{Tr}(\bullet)$, that outputs a fixed state $\tau$ irrespective of the input, is the erasure channel $\mathcal{P}_{\mathcal{N},\sigma}(\bullet)=\sigma\,\textrm{Tr}(\bullet)$ that re-prepares the prior $\sigma$. We will often omit the subscripts $\mathcal{N},\sigma$ from $\mathcal{P}_{\mathcal{N},\sigma}$ for brevity.

For our problem, we shall build on the following result in Ref.~\cite{GaussianPetz2}: given a Gaussian forward channel $\mathcal{N}$ characterized by $(X_\mathcal{N},Y_\mathcal{N},\mathbf{d}_\mathcal{N})$ and a Gaussian reference state $\sigma$ characterized by $(\overline{\mathbf{r}}_{\sigma},V_\sigma)$, the corresponding Petz recovery map $\mathcal{P}_{\mathcal{N},\sigma}$ is also a Gaussian channel, characterized by 
\begin{align}
    \begin{aligned}
        X_\mathcal{P} &= \left(I+(V_\sigma\Omega)^{-2}\right)^{\frac{1}{2}}V_\sigma X^T_\mathcal{N} \left(I+(\Omega V_{\mathcal{N(\sigma)}})^{-2}\right)^{-\frac{1}{2}}V_{\mathcal{N}(\sigma)}^{-1},\\
        Y_\mathcal{P} &= V_\sigma -X_\mathcal{P} V_{\mathcal{N(\sigma)}} X_\mathcal{P}^T, \,\,\text{and}\\
        \mathbf{d}_\mathcal{P} &= \overline{\mathbf{r}}_\sigma-X_\mathcal{P} \overline{\mathbf{r}}_\mathcal{N(\sigma)}. \label{eq:petz_XY}
    \end{aligned}
\end{align}

\section{\label{model}Lossy Channel and its Petz Recovery Map}

\begin{figure}[ht]
\centering\includegraphics[scale=0.22]{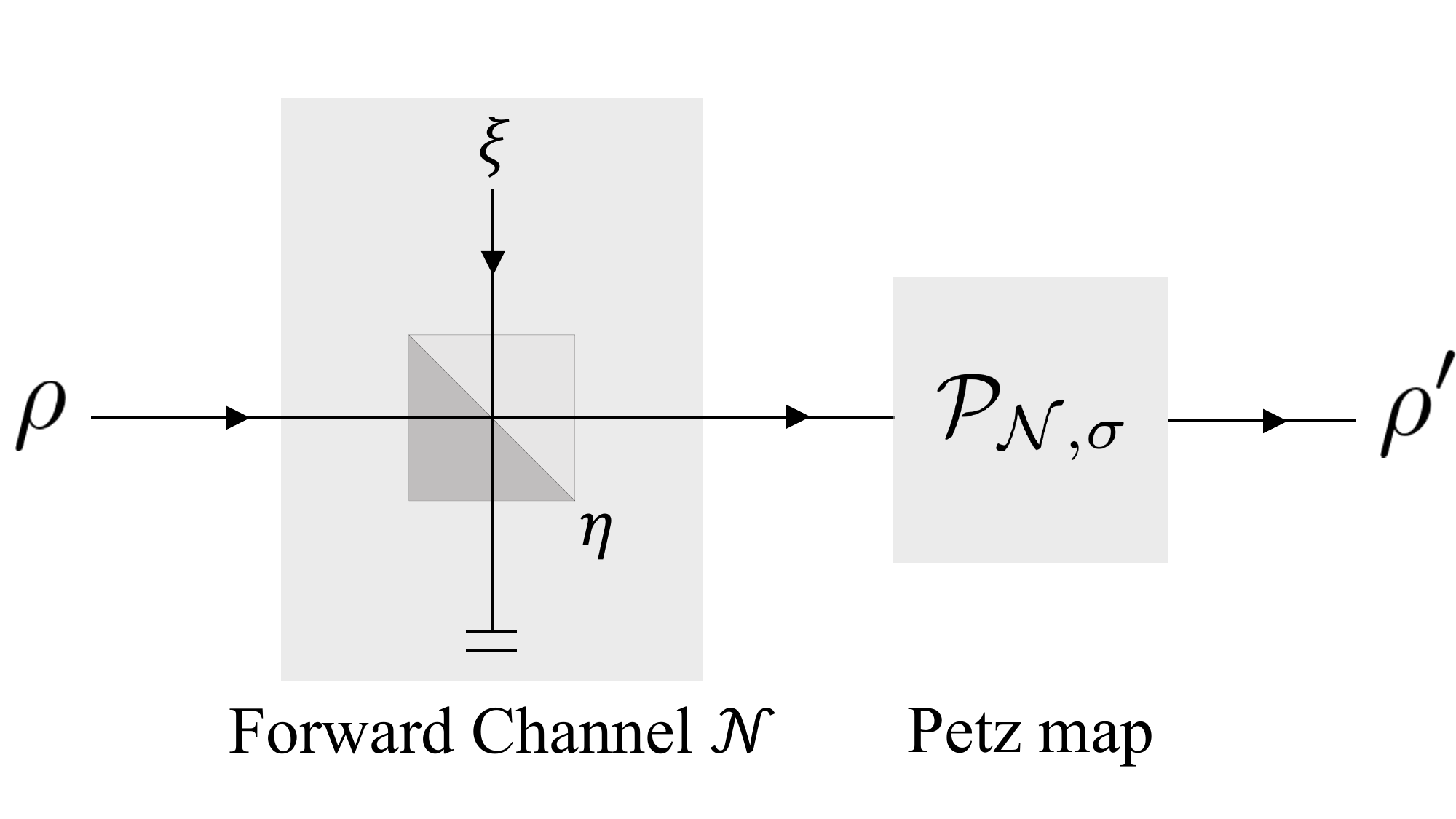}
\captionsetup{justification=raggedright}
    \caption{\label{fig:model}{\bf Setup of the noise model}. In the forward channel, a beam splitter with transmissivity $\eta$ is used and the environment in the state $\xi$ will be traced out after the beam splitter.}
\end{figure}

We focus now on the channel whose action we want to undo: state-independent losses in a single bosonic mode. To address a broad range of physical scenarios, including asymmetric or phase-sensitive noise, we model this lossy channel $\mathcal{N}$ as a single beam splitter with transmissivity $\eta$, where the mode under study is coupled to a general Gaussian environment $\xi$ (\cref{fig:model}) that is later discarded:
\begin{equation}
\mathcal{N}:\rho\rightarrow \text{tr}_{E}(U_{B}^{(\eta)}(\rho\otimes\xi_{E})U_{B}^{(\eta)\dagger}). \label{eq:loss_channel}
\end{equation}
Here, $U_{B}^{(\eta)}$ is the unitary operator of the beam splitter, a two-mode Gaussian unitary whose symplectic matrix is
\begin{equation}
X_{B}=\begin{bmatrix}\sqrt{\eta}I&\sqrt{1-\eta}I\\ -\sqrt{1-\eta}I&\sqrt{\eta}I\end{bmatrix} \label{eq:beamsplitter}
\end{equation}
where $\eta$ is the transmissivity of the beam splitter and $I$ is the $2\times2$ identity matrix. The Gaussian environment $\xi$ is characterized by a covariance matrix $V_{\xi}$ and a mean vector $\overline{\mathbf{r}}_{\xi}$. Thus, the Gaussian channel $\mathcal{N}$ can be expressed with:
\begin{equation}
X_{\mathcal{N}}=\sqrt{\eta}I, \quad Y_{\mathcal{N}}=(1-\eta)V_{\xi}, \quad d_{\mathcal{N}}=\sqrt{1-\eta}\overline{\mathbf{r}}_{\xi}. \label{eq:general_loss}
\end{equation}

If the prior $\sigma$ is also taken as Gaussian, the Petz recovery map is a Gaussian channel given by \cref{eq:petz_XY}. We focus on nontrivial lossy channels $0<\eta<1$. Plugging \cref{eq:general_loss} into \cref{eq:petz_XY} and using $M\Omega M^{T}=(\det M)\Omega$ for single-mode Gaussian states, one finds first:
\begin{equation}
X_{\mathcal{P}}=\left(\eta\frac{1-(\det V_{\sigma})^{-1}}{1-(\det V_{\mathcal{N}(\sigma)})^{-1}}\right)^{\frac{1}{2}}V_{\sigma}V_{\mathcal{N}(\sigma)}^{-1}.
\end{equation}
The Petz recovery map is another beam splitter under the conditions such that:
\begin{equation}
V_{\sigma}\propto V_{\mathcal{N}(\sigma)} \quad [\text{equivalently, } V_{\sigma}\propto V_{\xi}], \text{ and}
\end{equation}
\begin{equation}
\max\left(1, D_{\xi} - \Delta_{\xi}\right) \le \sqrt{\det V_{\sigma}} \le D_{\xi} + \Delta_{\xi},
\end{equation}
where we define $D_{\xi} \equiv \sqrt{\det V_{\xi}}$ and $\Delta_{\xi} \equiv \sqrt{(\det V_{\xi}-1)/\eta}$.

Specifically, under these conditions, the Petz recovery map $\mathcal{P}_{\mathcal{N},\sigma}$ can be realized by a beam splitter with transmissivity $\eta^{\prime}$ and a Gaussian ancilla $\xi^{\prime}$, characterized by $X_{\mathcal{P}}=\sqrt{\eta^{\prime}}I$, $Y_{\mathcal{P}}=(1-\eta^{\prime})V_{\xi^{\prime}}$, and $d_{\mathcal{P}}=\sqrt{1-\eta^{\prime}}\overline{\mathbf{r}}_{\xi^{\prime}}$, where $\eta^{\prime}=\eta^{\prime}(\sigma)$ is given by:
\begin{equation}
\eta^{\prime}=\eta\frac{\det V_{\sigma}-1}{\det V_{\mathcal{N}(\sigma)}-1} \label{eq:general_eta_prime}
\end{equation}
and $\xi^{\prime}=\xi^{\prime}(\sigma)$ is a Gaussian state $(V_{\xi^{\prime}}, \overline{\mathbf{r}}_{\xi^{\prime}})$ such that:
\begin{align}
    \begin{aligned}
        V_{\xi^{\prime}}&=\frac{1}{1-\eta^{\prime}}\left(1-\eta^{\prime}\sqrt{\frac{\det V_{\mathcal{N}(\sigma)}}{\det V_{\sigma}}}\right)V_{\sigma}, \\
        \overline{\mathbf{r}}_{\xi^{\prime}}&=\frac{1}{\sqrt{1-\eta^{\prime}}}(\overline{\mathbf{r}}_{\sigma}-\sqrt{\eta^{\prime}}(\sqrt{\eta}\overline{\mathbf{r}}_{\sigma}+\sqrt{1-\eta}\overline{\mathbf{r}}_{\xi})). 
    \end{aligned}
\end{align}
It can be verified that $\xi'$ is a physical state, i.e., $\det V_{\xi'} \geq 1$. 

\subsection{The Thermal Environment Limit}
In particular, when the environment $\xi$ and the reference state $\sigma$ are both thermal states, we reproduce the standard optical loss scenario where no phase coherence is expected between the environment and the input state. The thermal environment is expressed by its mean photon number $\overline{n}_{\xi}$, such that $V_{\xi}=(2\overline{n}_{\xi}+1)I$ and $\overline{\mathbf{r}}_{\xi}=0$. If the reference state has a mean photon number $\overline{n}_{\sigma}$, the generalized transmissivity in Eq. (\ref{eq:general_eta_prime}) takes the simple form:
\begin{equation}
\eta^{\prime}=\eta\frac{(2\overline{n}_{\sigma}+1)^{2}-1}{(\eta(2\overline{n}_{\sigma}+1)+(1-\eta)(2\overline{n}_{\xi}+1))^{2}-1}. \label{eq:eta'}
\end{equation}
While certainly $\eta^{\prime}\ge0$, it is not guaranteed that $\eta^{\prime}\le1$.

\begin{rst}
    \label{result1}
    Let $\mathcal{N}$ be the lossy channel with $0<\eta<1$ and thermal ancilla $\xi$. Its Petz recovery map $\mathcal{P}_{\mathcal{N},\sigma}$ for a thermal reference state $\sigma$ is implementable by a beam splitter of a transmittivity $0\le\eta^{\prime}(\sigma)<1$ and an ancilla state $\xi^{\prime}(\sigma)$, if $\sigma$ is such that:
    \begin{equation}
    0\le\overline{n}_{\sigma}<\overline{n}_{\xi}+\sqrt{\frac{\overline{n}_{\xi}(\overline{n}_{\xi}+1)}{\eta}}. \label{eq:result1_bound}
    \end{equation}
\end{rst}
When this is the case, the added noise matrix $Y_{\mathcal{P}}$ can be written as $Y_{\mathcal{P}}=(1-\eta^{\prime})V_{\xi^{\prime}}$, where $V_{\xi^{\prime}}$ is the covariance matrix of another thermal ancilla $\xi^{\prime}$ with mean photon number
    \begin{equation}
    \overline{n}_{\xi^{\prime}}=\frac{1}{1-\eta^{\prime}}((1-\eta\eta^{\prime})(2\overline{n}_{\sigma}+1)-(1-\eta)\eta^{\prime}(2\overline{n}_{\xi}+1)).
    \end{equation}
Indeed, $\overline{n}_{\xi^{\prime}}\ge0$ whenever $0\le\eta^{\prime}<1$ is satisfied.

When \cref{result1} does not hold and $\eta^{\prime}=1$, the Petz recovery map is an additive noise channel with noise matrix $Y_{\mathcal{P}}\ge0$; or when $\eta^{\prime}>1$, it is a phase-insensitive amplifier, with $\eta^{\prime}$ corresponding to the amplification gain. Henceforth, we will call $\eta^{\prime}$ a generalized transmissivity, which can take values in range $[0,\infty)$. See Appendix~\ref{general solution} for the proof of \cref{result1}.

Now, in many realistic cases, notably optical frequencies, the thermal environment of the forward channel is the vacuum to an excellent degree of approximation. It then follows from the condition (\ref{eq:result1_bound}) that the Petz map is a beam splitter only when $\overline{n}_{\sigma}=0$, when the thermal prior is also chosen to be the vacuum (i.e., $\overline{n}_{\xi}=0$). For all other thermal reference states, the Petz map acts as a phase-insensitive amplifier. It is intriguing that the Petz recovery map can, and often is, a phase-insensitive amplifier: qualitatively, it can be seen as trying to amplify the information left in the signal after the losses, rather than just trying to recover the average number of photons.

\section{\label{fidelity}Recovery performance of the Petz Recovery Map}

\begin{figure*}
    \centering
    \includegraphics[width=1\linewidth]{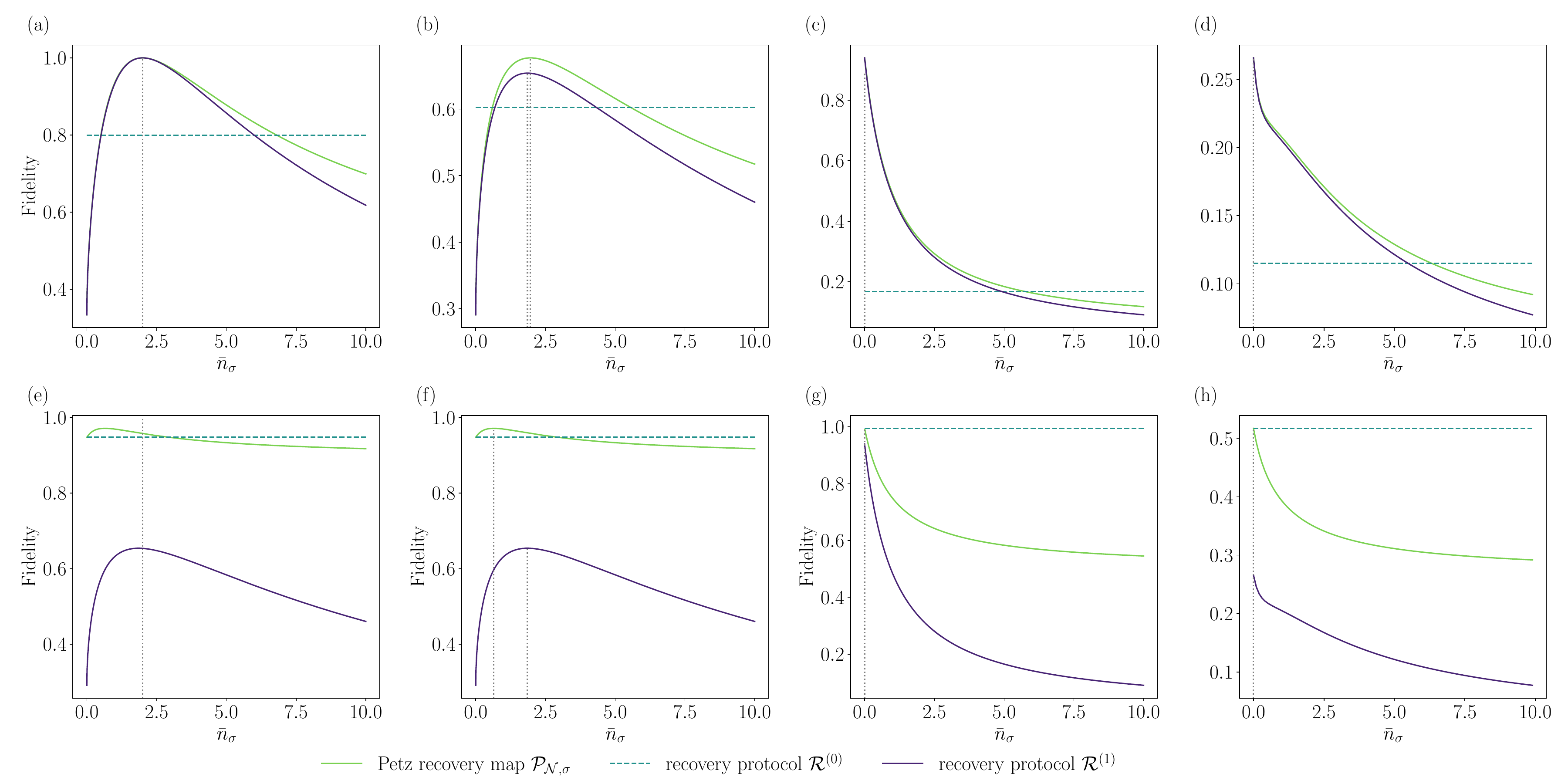}
    \captionsetup{justification=raggedright}
    \caption{{\bf Comparison between the Petz recovery $\mathcal{P}_{\mathcal{N},\sigma}$ and the trivial recovery protocols $\mathcal{R}^{(0)},\mathcal{R}^{(1)}_{\sigma}$.} A forward noise model considered here is a beam splitter of transmissivity $\eta=0.5$. Subfigures (a)–(d) correspond to the case where the environment of the forward channel is a thermal state with mean photon number $\overline{n}_\xi = 10$. In this regime, the Petz recovery map operates as a beam splitter over $0 \leq \overline{n}_\sigma \leq 10$. Subfigures (e)–(h) correspond to the case where the environment of the forward channel is the vacuum state, i.e., $\overline{n}_\xi = 0$. In this case, the Petz recovery map operates as a phase insensitive amplifier over $0 \leq \overline{n}_\sigma \leq 10$. We consider a reference state $\sigma$ that is also a thermal state 
    with its photon number $\overline{n}_\sigma$.
    The figure shows the plots of the fidelity against $\overline{n}_\sigma$ between a input state and a state recovered by $\mathcal{P}_{\mathcal{N},\sigma}$ (green line), $\mathcal{R}^{(1)}_\sigma$ (purple line), and $\mathcal{R}^{(0)}$ (dashed blue line), while considering three input states: (a) and (e) thermal state  $V_{th}=(2\overline{n}_{th}+1)I$ with mean photon number $\overline{n}_{th}=2$, (b) and (f) a  squeezed state $V_{sq} = \rm{diag}(2.5,10)$ with  $\overline{\textbf{r}}_{sq} =0$,
    (c) and (g) a coherent state $\ket{\alpha}$ with amplitude $\alpha = 1/2\sqrt{2}(1+i)$, (d) and (h) a even cat state with $\ket{\alpha}$ with amplitude $\alpha = 1+i$. In all plots, the dotted line indicates when respective recovery protocols achieves the optimal fidelity.
    } 
    \label{fig:fidelity fig}
\end{figure*}

In this section we study the performance of the Petz recovery map for input states different from the reference state. It is trivial that $\mathcal{P}_{\mathcal{N},\sigma}$ is realizable with a beam splitter when $\eta'=1$ (identity channel: the Petz map is also a beam splitter with $\eta=1$) or $\eta=0$ (erasure channel: the Petz map is also a beam-splitter $\eta'=0$ with $\xi'=\sigma$). Let us then focus on nontrivial lossy channels $0<\eta<1$. We consider the simple case where the environment and reference state are thermal states. 

\subsection{Comparison between passive recovery protocols}

When one considers a recovery protocol, the following protocols set natural benchmarks:
\begin{enumerate}
    \item A recovery protocol $\mathcal{R}^{(0)}$ consisting in keeping the noisy state doing nothing. This is the Petz recovery map for the identity channel, i.e.,~for $\eta=1$.
    \item A recovery protocol $\mathcal{R}^{(1)}_{\sigma}$ consisting in discarding the noisy state and replacing it with one's belief state $\sigma$. This is the Petz recovery map for the erasure channel, i.e.~for $\eta=0$.
\end{enumerate}

We will make use of the fidelity~\cite{uhlmann1976fidelity,jozsa1994fidelity} between the initial state and the recovered state to compare performance of these recovery protocols with that of Petz recovery maps. The fidelity $F(\rho_1,\rho_2)$ between $\rho_1$ and $\rho_2$ is a faithful measure of how close $\rho_1$ and $\rho_2$, defined as $F(\rho_1,\rho_2)\equiv \left(\Tr\sqrt{\sqrt{\rho_2}\rho_1\sqrt{\rho_2}}\right)^2$. For example, $F(\rho_1,\rho_2)=1$ if $\rho_1=\rho_2$, and $F(\rho_1,\rho_2)=0$ for orthogonal states $\rho_1,\rho_2$.

We show that it depends on a reference state whether Petz recovery maps are better than $\mathcal{R}^{(0)}$ while Petz recovery maps are always better than $\mathcal{R}^{(1)}_{\sigma}$ when considering thermal input state. The following result presents the condition of a reference state when Petz recovery maps outperform $\mathcal{R}^{(0)}$ (see Appendix~\ref{app:fidelity0} for the proof):

\begin{rst}
For any thermal states $\rho$, if $\max(1, \min (z_0,z_1)) \le g(\sigma) \leq \max (z_0, z_1)$, then
    \begin{align}
        F(\rho, \mathcal{R}^{(0)}\circ\mathcal{N}(\rho)) \leq F(\rho, \mathcal{P}_{\mathcal{N},\sigma}\circ\mathcal{N}(\rho))\,;
    \end{align}
    where we define
    \begin{align}
        g(\sigma)&:= \eta\eta'(\sigma)(2\overline{n}_\rho+1) + \left(1-\eta\eta'(\sigma)\right)(2\overline{n}_\sigma+1)\\
        z_0 &:= \eta (2\overline{n}_\rho+1)+ (1-\eta)(2\overline{n}_\xi+1)\\
        z_1 &:= 2(2f(z_0)-1)(2\overline{n}_\rho+1)-z_0\\
        f(z)&:= \frac{1}{2}\left( z(2\overline{n}_\rho+1)+1 - 2\sqrt{\overline{n}_\rho(\overline{n}_\rho+1)(z^2-1)} \right).\nonumber
    \end{align}
\end{rst}

This implies that if one's guess on the actual initial state is significantly incorrect so that the reference state is considerably different from the initial state, then it is better not to carry out the Petz recovery maps but leave the noisy state as it is. 

The next result shows that the Petz recovery maps always outperform the protocol $\mathcal{R}^{(1)}_\sigma$ which simply replaces the noisy state with one's belief state $\sigma$ in the same scenario (see Appendix~\ref{app:fidelity1} for the proofs):

\begin{rst}
For any thermal input states $\rho$, it always has
    \begin{align}
        F(\rho, \mathcal{R}^{(1)}_\sigma\circ\mathcal{N}(\rho)) \leq F(\rho, \mathcal{P}_{\mathcal{N},\sigma}\circ\mathcal{N}(\rho)).
    \end{align}
\end{rst}
It is noteworthy that these two results hold beyond the case of thermal inputs. Although we do not know if they hold in general, in the examples that we tested numerically we have always found these relations satisfied (see e.g. Fig.~\ref{fig:fidelity fig}). In the same figure, we further observe that choosing a reference state closer to the initial state generally improves the recovery performance. However, fidelity is not the appropriate metric in this context, as the closest reference state to the actual input state $\rho$ with respect to fidelity, that is, the reference state $\sigma$ with which $\mathcal{R}^{(1)}_\sigma$ achieves the maximum fidelity between the input $\rho$ and the recovered state (in this case $\sigma$), does not necessarily yield the optimal Petz recovery, as seen in Fig.~2(b) and Fig.~2(e). 

\subsection{Near-optimality of the Petz map among recovery maps}

\begin{figure*}
    \centering
    \includegraphics[width=1\linewidth]{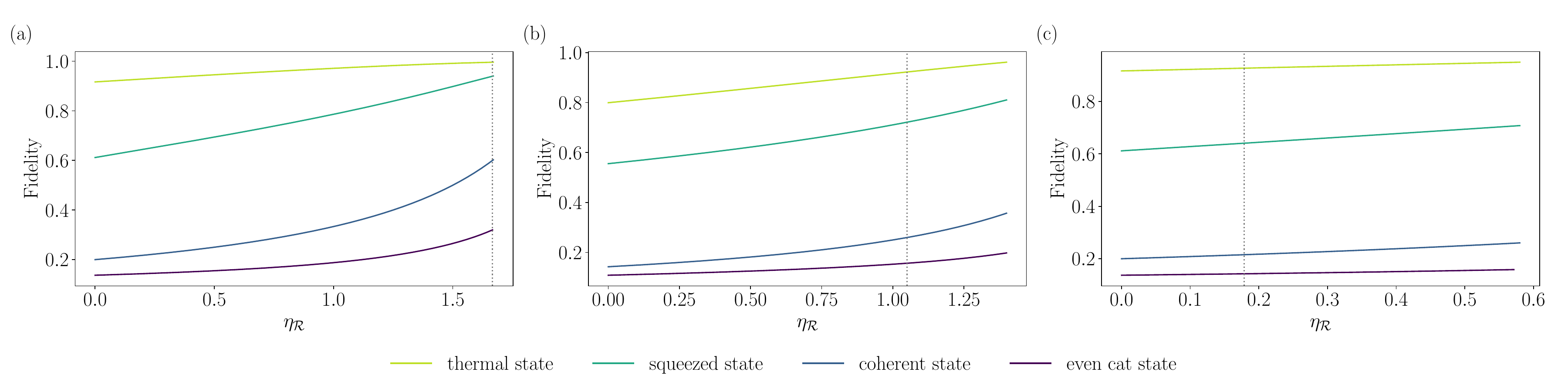}
    \caption{\textbf{Comparison between the Petz recovery $\mathcal{P}_{\mathcal{N},\sigma}$ and and other recovery protocols $\mathcal{R}\in \mathfrak{R}_{\mathcal{N},\sigma}$.} 
    The transmissivity of the forward channel is set to be $\eta=0.5$. The dotted line is a guide for the eye to show where the transmissivity of Petz recovery map lies. The mean photon numbers of the environment and the mean photon numbers of the thermal reference states are (a) $\overline{n}_\xi = 0$ and $\overline{n}_\sigma = 4$, (b) $\overline{n}_\xi = 2$ and $\overline{n}_\sigma = 6$, and (c) $\overline{n}_\xi = 10$ and $\overline{n}_\sigma = 4$, respectively. Notably, (a) and (b) are the cases where the Petz recovery map is a phase-insensitive amplifier while in (c), it is a beam splitter. The four input state are considered in all of these cases; thermal state $V_{th}=(2\overline{n}_{th}+1)I$ with mean photon number $\overline{n}_{th}=2$; a squeezed state $V_{sq} = \rm{diag}(2.5,10)$ with  $\overline{\textbf{r}}_{sq} =0$;
    a coherent state $\ket{\alpha}$ with amplitude $\alpha = 1/2\sqrt{2}(1+i)$; and a even cat state with amplitude $\alpha = 1+i$.  }
    \label{fig:comparison1}
\end{figure*}

In order to further appreciate the specificity of the Petz map, we take a step back and observe the following. If the goal were merely to recover the reference thermal state $\sigma$ after the lossy channel $\mathcal{N}$, any recovery channel  $(X_\mathcal{R}, Y_\mathcal{R}, \mathbf{d}_\mathcal{R})$ would do it that satisfies $X_\mathcal{R}V_{\mathcal{N}(\sigma)}X_\mathcal{R}^T + Y_\mathcal{R} = V_\sigma$ and $X_\mathcal{R}\overline{\mathbf{r}}_{\mathcal{N}(\sigma)} + \mathbf{d}_\mathcal{R} = \overline{\mathbf{r}}_{\sigma}$, alongside with the CP condition~\eqref{eq:CP}. Since the scenario under study is the case where the lossy channel has $X_{\mathcal{N}}\propto I$ and the thermal prior has also $V_{\sigma}\propto I$ and $\mathbf{d}_\sigma=0$, we thus restrict our attention to the set $\mathfrak{R}_{\mathcal{N},\sigma}$ of the recovery protocols $(X_\mathcal{R} = \sqrt{\eta_\mathcal{R}}I, Y_\mathcal{R}, \mathbf{d}_\mathcal{R} = 0)$ with $\eta_\mathcal{R}\ge0$ satisfying 
\begin{align}
        &X_\mathcal{R}V_{\mathcal{N}(\sigma)}X_\mathcal{R}^T + Y_\mathcal{R} = V_\sigma,\\
        &X_\mathcal{R}\overline{\mathbf{r}}_{\mathcal{N}(\sigma)} = \overline{\mathbf{r}}_{\sigma},\\
        &Y_\mathcal{R} + i\Omega \geq i X_\mathcal{R} \Omega X_\mathcal{R}^T. 
\end{align} 

In what follows, we will use $(\eta_\mathcal{R},Y_\mathcal{R})$ in place of $(X_\mathcal{R} = \sqrt{\eta_\mathcal{R}}I, Y_\mathcal{R}, \mathbf{d}_\mathcal{R} = 0)$. The Petz recovery map is one such channel $(\eta_\mathcal{P}=\eta', Y_\mathcal{P})$ with $\eta'$ given in \cref{eq:eta'}. We are going to gain some insight into why the Petz recovery map selects that specific value.

First notice that the CP condition \eqref{eq:CP} bounds the generalized transmissivity as
\begin{align}
    \eta_\mathcal{R} \;\leq\;
    \begin{cases}
        \min (1,\dfrac{\overline{n}_\sigma}{\eta\, \overline{n}_\sigma + (1-\eta)\, \overline{n}_\xi}), & 0 \leq \eta_\mathcal{R} \leq 1, \\[2.0ex]
        \max(1,\dfrac{\overline{n}_\sigma + 1}{\eta\, \overline{n}_\sigma + (1-\eta)\, \overline{n}_\xi + 1}), & \eta_\mathcal{R} > 1.
    \end{cases}\label{etaRbounds}
\end{align}
In particular, when the environment is the vacuum state, $\overline{n}_\xi = 0$, while the reference state is not the vacuum state, $\overline{n}_\sigma > 0$, the Petz recovery map is a phase-insensitive amplifier, i.e., $\eta_\mathcal{P}>1$ and corresponds to the recovery channel that saturates the CP bound (the second inequality in the CP condition~\eqref{etaRbounds}). In this case, rather than just fully recovering the reference state, the Petz recovery map preserves the largest information from the input by choosing the largest $\eta_\mathcal{R}$ and introducing the least amount of noise. When the environment is not the vacuum state, although the Petz recovery map does not saturate the bound, it still features a relatively large $\eta_\mathcal{R}$ that remains close to the optimal limit, see \cref{fig:comparison1}.

The inequalities also indicate that, for any given reference thermal state and a thermal loss channel, there always exists non-trivial beam splitters that can perfectly recover the reference state, in addition to $\mathcal{R}^{(0)}$ and $\mathcal{R}^{(1)}_\sigma$. Among these recovery protocols, the Petz recovery map shows a near-optimal performance. \cref{fig:comparison1} shows numerical examples of this argument. In particular, Fig.~3(a) is the special case when the environment of the forward channel is the vacuum. The Petz recovery map is the optimal one among these recovery channels in $\mathfrak{R}_{\mathcal{N},\sigma}$. Narrowing our focus to recovery protocols implemented via beam splitters for comparison with Petz recovery maps, we find that when $\eta_\mathcal{P} >1$, the Petz recovery map has a better recovery performance than using another beam splitter, as seen in Fig.~3(a) and Fig.~3(b). In contrast, when $0 \leq \eta_\mathcal{P} \leq 1$, there exists recovery protocols implemented via beam splitters that perform better recovery than the Petz recovery map, as seen in Fig.~3(c).

\begin{figure*}
    \centering
    \includegraphics[width=0.7\linewidth]{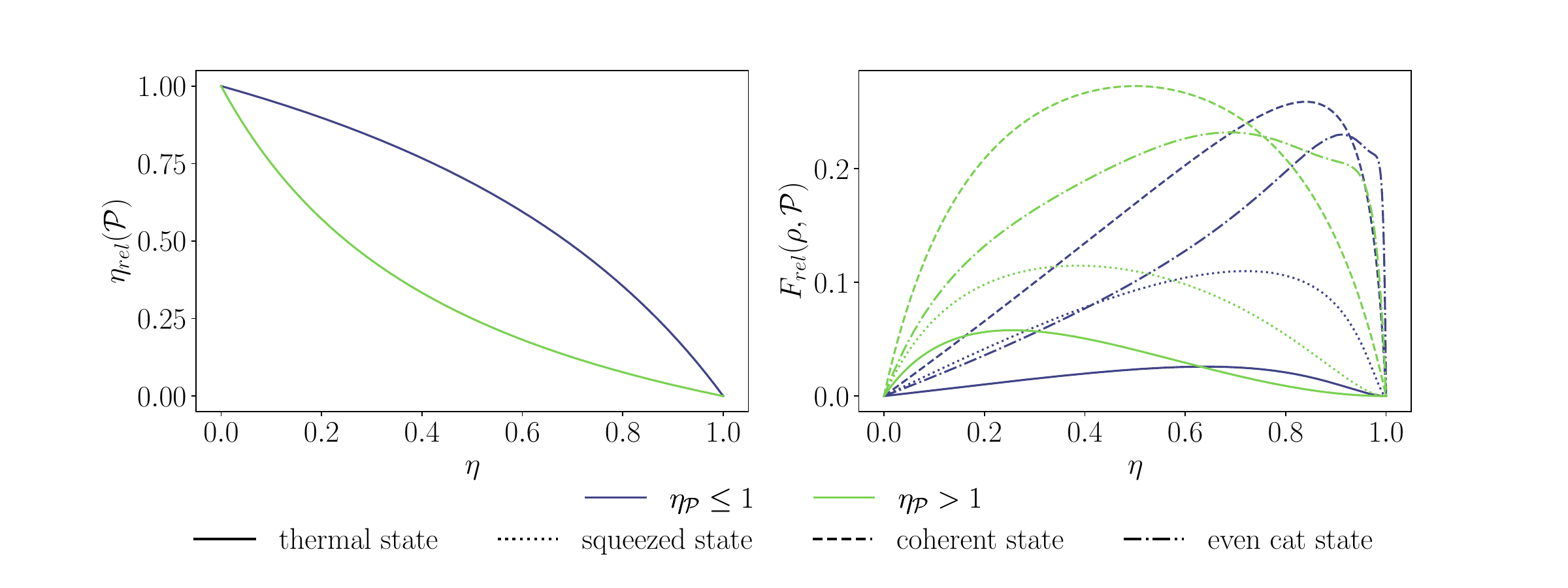}
    \caption{\textbf{The relative transmissivity difference $\eta_{\text{rel}}(\mathcal{P})$ and the relative fidelity difference $F_{\text{rel}}(\rho, \mathcal{P})$ against the transmissivity $\eta$ of the forward channel $\mathcal{N}$.} The purple line is the case where the environment of the forward channel is a thermal state with $\overline{n}_\xi = 10$ and the reference state is a thermal state with $\overline{n}_\sigma = 4$, such that $\eta_\mathcal{P} \leq 1$. The green line is the case where the environment of the forward channel is a thermal state with $\overline{n}_\xi = 2$ and the reference state is another thermal state with $\overline{n}_\sigma = 6$, such that $\eta_\mathcal{P} >1$. Here, we consider three input states; thermal state $V_{th}=(2\overline{n}_{th}+1)I$ with mean photon number $\overline{n}_{th}=2$ (solid line); a squeezed state $V_{sq} = \rm{diag}(2.5,10)$ with  $\overline{\textbf{r}}_{sq} =0$ (dotted line); a coherent state $\ket{\alpha}$ with amplitude $\alpha = 1/2\sqrt{2}(1+i)$ (dashed line); and a even cat state with $\alpha = 1+j$ (dashdoted line). 
    }
    \label{fig:relative difference}
\end{figure*}

Furthermore, we also compare the difference between the Petz recovery map and the optimal recovery protocol $\mathcal{R}_{\text{op}}$ in $\mathfrak{R}_{\mathcal{N},\sigma}$, shown in \cref{fig:relative difference}. Here we introduce the \emph{relative transmissivity difference} between the generalized transmissivity of the Petz recovery map and the maximum generalized transmissivity defined as 
\begin{align}
    \begin{aligned}
        \eta_{\text{rel}}(\mathcal{P}) \equiv \frac{\eta_{\max}-\eta_\mathcal{P}}{\eta_{\max}}
    \end{aligned}
\end{align}
with $\eta_{\max} \equiv \max \{ \eta_\mathcal{R}: \mathcal{R} \in \mathfrak{R}_{\mathcal{N},\sigma}\}$. Additionally, we introduce the \emph{relative fidelity difference} between the Petz recovery map and 
the optimal recovery protocol, defined as
\begin{align}
    \begin{aligned}
        F_{\text{rel}}(\rho,\mathcal{P}) &\equiv \frac{F_{\max}(\rho, \mathcal{N})-F(\rho,\mathcal{P}\circ \mathcal{N}(\rho))}{F_{\max}(\rho,\mathcal{N})}, 
    \end{aligned}
\end{align}
with $F_{\max}(\rho, \mathcal{N}) \equiv \max_{\mathcal{R}\in \mathfrak{R_{\mathcal{N},\sigma}}} F(\rho,\mathcal{R}\circ \mathcal{N}(\rho))$.

In particular, the optimal recovery protocol $\mathcal{R}_{\text{op}}$ is the one that reaches the maximum fidelity, i.e., $F_{\max}(\rho, \mathcal{N}) = F(\rho,\mathcal{R}_\text{op}\circ \mathcal{N}(\rho)).$
Although we suppress the argument $\mathcal{N}$ for brevity, $F_{\text{rel}}(\rho,\mathcal{P})$ depends on $\mathcal{N}$, thus on $\eta$.

From \cref{fig:relative difference}, we make an observation that the less lossy the forward channel is, the closer the Petz map approaches the optimal performance. Such behavior has been proven in various contexts using different optimality criteria~\cite{nearoptimal1,nearpotimal2}. Therefore this observation aligns with our expectation that Petz map gets closer to the optimal recovery when the noise is closer to the perfectly correctable cases. Although not strictly optimal, the relative fidelity difference, compared to the optimal recovery channel is below $25\%$, and drops below $15\%$ when the input state shares the same structure as the reference state—in our case, the mean vectors of input states are zero, $\mathbf{\overline{\mathbf{r}}} = 0$. Collectively, these results show the near-optimality of Petz recovery maps in the scenario of losses in an optical mode. 

\begin{figure*}
    \centering
    \includegraphics[width=0.8\linewidth]{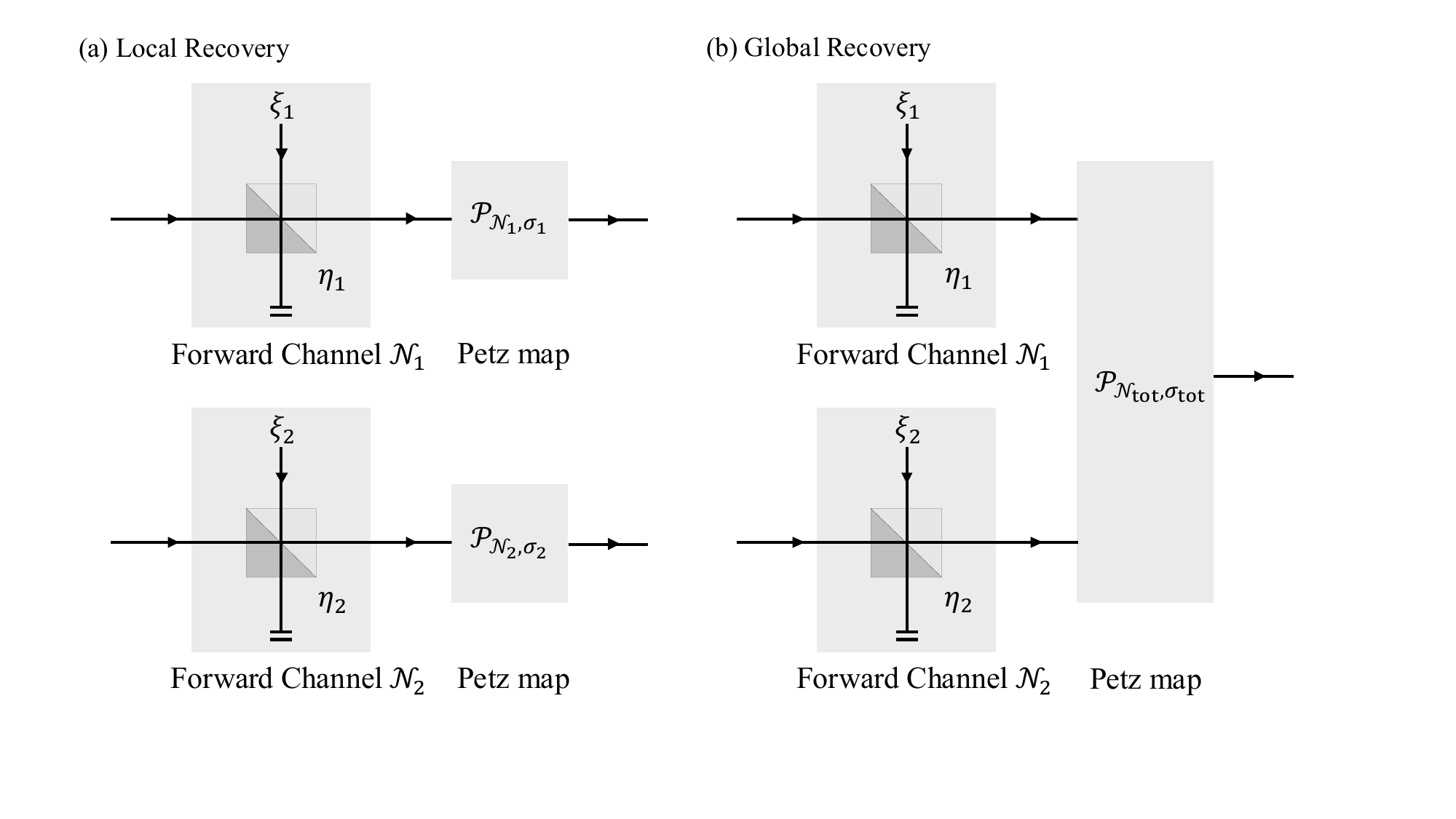}
    \caption{\textbf{Setup of two recoveries.} (a) Local recovery: The two modes pass through independent forward loss channels ($\mathcal{N}_1$ and $\mathcal{N}_2$), modeled as beam splitters interacting with environmental states $\xi_1$ and $\xi_2$. The recovery is performed independently on each mode using local Petz recovery maps ($\mathcal{P}_{\mathcal{N}_1,\sigma_1}$ and $\mathcal{P}_{\mathcal{N}_2,\sigma_2}$), corresponding to a separable reference state. (b) Global recovery: The modes undergo the same independent forward loss channels, but the recovery is executed via a single, joint two-mode Petz recovery map ($\mathcal{P}_{\mathcal{N}_\text{tot},\sigma_\text{tot}}$). This global approach utilizes a non-separable  reference state to exploit shared quantum correlations during the recovery process. }
    \label{fig:multi}
\end{figure*}

\begin{figure*}
    \centering
    \includegraphics[width=1\linewidth]{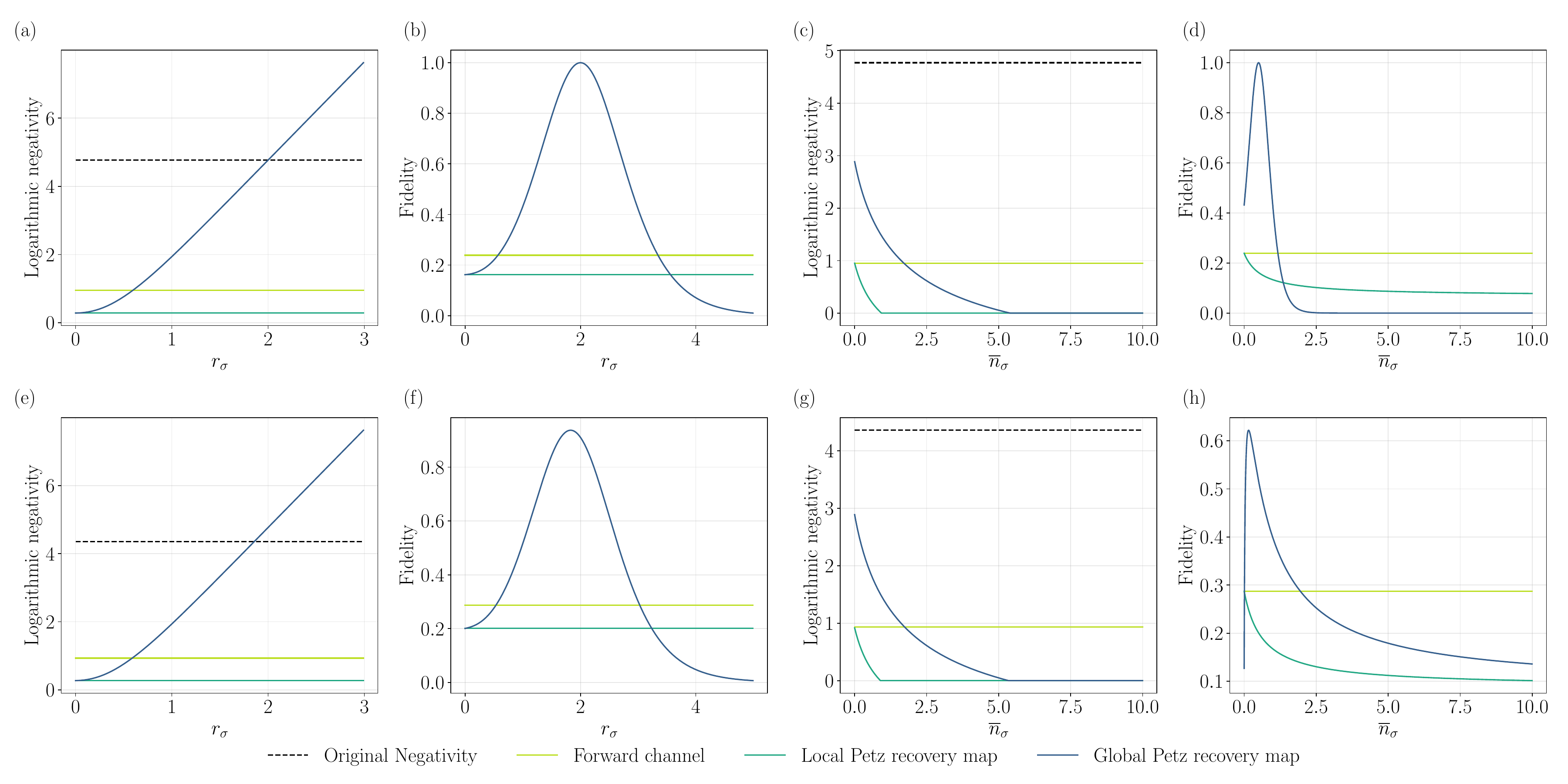}
    \caption{\textbf{Comparison of local and global Petz recovery maps.} For all eight figures, the  forward channel is set to be identical beam splitters with transmissivity $\eta = 1/2$ and two identical vacuum environment. In (a)-(d), the input state is a symmetric squeezed thermal state with mean photon number of each mode $\overline{n}_{in} = 1/2$ and squeezing parameter $r_{in} = 2$ and in (e)-(h), the input state is squeezed thermal state with mean photon number of two mode $\overline{n}_1 = 1/2, \overline{n}_2 = 3/2$ and squeezing parameter $r_{in} = 2$. For left four figures, the reference state of the local Petz recovery map is set to be symmetric thermal state with mean photon number $\overline{n}_\sigma = 1/2$ and the reference state of the global Petz recovery map is set to be symmetric squeezed thermal state with mean photon number of each mode $\overline{n}_\sigma = 1/2$. For the right four figures, the the reference state of the local Petz recovery map is set to be symmetric thermal state and the reference state of the global Petz recovery map is set to be symmetric squeezed thermal state with squeezing parameter $r_\sigma = 1$. The black dotted line represents the logarithmic negativity of the input state, the green line represents the logarithmic negativity or fidelity after the forward channel, the blue line represents the logarithmic negativity or fidelity after the local Petz recovery map, and the purple line represents the logarithmic negativity or fidelity after the global Petz recovery map.}
    \label{fig:comparison2}
\end{figure*}

\section{\label{multi}EXTENSION TO TWO-MODE LOSSY CHANNELS}

The preceding sections focused on single-mode optical losses: it is natural to extend the study to multi-mode systems. We do not attempt an exhaustive study, but rather focus on a specific example and question: we consider a two-mode Gaussian loss channel which is characterized by two beam splitters; and we study the advantage of performing a joint two-mode recovery map (``global recovery''), as compared to performing on each mode separately the recovery discussed in the previous sections (``local recovery'').

\subsection{Two-Mode Beam Splitter and Recovery Frameworks}

To extend our analysis, we model the two-mode loss as a pair of beam splitters interacting with a Gaussian environment. There are two ways to treat the recovery of such a channel: a local recovery and a global recovery, see \cref{fig:multi}. 

In a local recovery scheme, the modes are treated independently using a separable reference state, e.g., $\sigma_\text{tot} = \sigma_1 \otimes \sigma_2$. This approach factorizes the Petz recovery map into independent, single-mode operations, i.e., $\mathcal{P}_{\mathcal{N}_\text{tot},\sigma} = \mathcal{P}_{\mathcal{N}_1,\sigma_1} \otimes \mathcal{P}_{\mathcal{N}_2,\sigma_2}$. Conversely, a global recovery scheme utilizes a non-separable reference state, such as an entangled two-mode squeezed thermal state, to leverage shared correlations during the recovery process. 

The forward Gaussian channel $\mathcal{N}_\text{tot}$ can be expressed with
\begin{align}
   \begin{aligned}
        X_{\mathcal{N}_\text{tot}} &= \sqrt{\eta_1} I \oplus \sqrt{\eta_2}I, \\
        Y_{\mathcal{N}_\text{tot}} &= (1-\eta_1)V_{\xi_1} \oplus (1-\eta_2)V_{\xi_2}, \\
        \mathbf{d}_{\mathcal{N}_\text{tot}} &= \sqrt{1-\eta_1} \overline{\mathbf{r}}_{\xi_1} \oplus \sqrt{1-\eta_2} \overline{\mathbf{r}}_{\xi_2},
   \end{aligned}
\end{align}
where $\eta_{1,2}$ are the transmissivities of the corresponding beam splitters, $V_{\xi_{1,2}}$ are the covariance matrices of the corresponding environment, and $\overline{\mathbf{r}}_{\xi_{1,2}}$ are the corresponding mean vector of the environment. The set up of these two recoveries are shown in \cref{fig:multi-mode model}. 

For local recovery, the Petz recovery map is simply applying two single-mode Petz recovery map locally. For a general non-separable Gaussian reference and possibly different lossy channels, the global Petz map is complex. But there is a natural example, in which it takes a clear form: 

\begin{rst}
    \label{result:two_mode_implementability}
    Let $\mathcal{N}_{tot} = \mathcal{N}_1 \otimes \mathcal{N}_2$ be a symmetric two-mode lossy channel with identical transmissivity $0<\eta_1=\eta_2=\eta<1$ and a symmetric thermal ancilla $\xi$ with mean photon number $\overline{n}_{\xi}$; and let the reference state be the symmetric two-mode squeezed thermal reference state $\sigma$, characterized by mean photon number $\overline{n}_{\sigma}$ and real squeezing parameter $r_{\sigma}$. The Petz recovery map $\mathcal{P}_{\mathcal{N},\sigma}$ is implementable by a combination of a two-mode squeezer $S(r^{\prime})$ and a pair of beam splitters or phase-insensitive amplifier of generalized transmittivity $\eta'$ given by
    \begin{align}
        \begin{aligned}
            X_\mathcal{P} &= \sqrt{\eta \frac{1-(\det V_\sigma)^{-1/2}}{1-(\det V_{\mathcal{N}(\sigma)})^{-1/2}}} V_\sigma V_{\mathcal{N}(\sigma)}^{-1}\\
            &\equiv X(\eta') S(r'),
        \end{aligned}
    \end{align}
    where $X(\eta')$ is the transformation matrix of beam splitter or phase-insensitive amplifier with $\eta' = \eta \frac{1-(\det V_\sigma)^{-1/2}}{1-(\det V_{\mathcal{N}(\sigma)})^{-1/2}} \frac{(\det V_\sigma)^{1/2}}{(\det V_{\mathcal{N}(\sigma)})^{1/2}}$, and the squeezing parameter $r'  = \arctanh \left (\frac{(1-\eta)(2\overline{n}_\xi+1)}{\eta(2\overline{n}_\sigma+1)+(1-\eta)(2\overline{n}_\xi+1)} \tanh (2r_\sigma)
    \right )$. 
\end{rst}
The proof is given in Appendix \ref{two mode}.

\subsection{Recovery of Quantum entanglement}

For multi-mode continuous-variable systems, it is important to consider not only the overall state fidelity but also the recovery of quantum correlations, such as entanglement. Losses inevitably degrade these inter-mode correlations, and it is interesting to know how much difference exists between a local recovery and a global recovery in restoring them. Generally, global recovery schemes outperform local recovery schemes in preserving and restoring entanglement, as the entangled reference state introduces correlation-preserving structures directly into the recovery channel, shown in \cref{fig:comparison2}. Here we use logarithmic negativity\cite{vidal2002computable, adesso2004extremal} as a measure of entanglement. It is worth noting that while the reference state is squeezed, it exhibits zero negativity in the regime where the squeezing parameter is $\abs{r_\sigma} \leq \frac{1}{2} \ln (\overline{n}_\sigma + 1)$, with $\overline{n}_\sigma$ denoting the mean photon number per mode. Despite this zero negativity, the global Petz recovery map still outperforms the local Petz recovery map in recovering correlations. However, as the mean photon number $\overline{n}_\sigma$ becomes sufficiently large, the performance gap between the two maps narrows, indicating that the local Petz recovery map can achieve results comparable to the global one. Furthermore, as either the squeezing parameter $r_\sigma$ or the mean photon number $\overline{n}_\sigma$ increases, the overall state fidelity recovered by the global map can degrade. Consequently, selecting these parameters requires a careful trade-off to effectively balance correlation recovery against state fidelity.

\section{Conclusion}
We have studied the Petz recovery map of optical losses, for thermal reference states. We have found that, depending on the choice of the reference state, the Petz recovery map is either another lossy channel with suitable ancilla, or a phase-independent amplifier.

We have further showed that the Petz recovery map performs at least as well as, and often better than, a recovery protocol that replaces the noisy state with a reference state. We have also identified conditions under which applying the Petz map may be less effective than simply retaining the noisy state. We show that the Petz recovery map is truly near-optimal when comparing to a class of recovery maps. 

Finally, we have extended our analysis to the multi-mode lossy channel. Though a global Petz recovery map is generally superior at recovering entanglement, there are parameter regimes where the local Petz map achieves comparable performance. Besides, a careful parameter trade-off is required to effectively balance correlation recovery against overall state fidelity.

\begin{acknowledgments}
We thank Alexia Auff\`eves for the question that triggered this work, and Ge Bai, Lin Htoo Zaw, Mingxuan Liu and Wenhan Png for helpful discussions. This project is supported by the National Research Foundation, Singapore through the National Quantum Office, hosted in A*STAR, under its Centre for Quantum Technologies Funding Initiative (S24Q2d0009); and by the Ministry of Education, Singapore, under the Tier 2 grant ``Bayesian approach to irreversibility'' (Grant No.~MOE-T2EP50123-0002).
\end{acknowledgments}

\bibliography{reference}

\begin{thebibliography}{35}%
\makeatletter
\providecommand \@ifxundefined [1]{%
 \@ifx{#1\undefined}
}%
\providecommand \@ifnum [1]{%
 \ifnum #1\expandafter \@firstoftwo
 \else \expandafter \@secondoftwo
 \fi
}%
\providecommand \@ifx [1]{%
 \ifx #1\expandafter \@firstoftwo
 \else \expandafter \@secondoftwo
 \fi
}%
\providecommand \natexlab [1]{#1}%
\providecommand \enquote  [1]{``#1''}%
\providecommand \bibnamefont  [1]{#1}%
\providecommand \bibfnamefont [1]{#1}%
\providecommand \citenamefont [1]{#1}%
\providecommand \href@noop [0]{\@secondoftwo}%
\providecommand \href [0]{\begingroup \@sanitize@url \@href}%
\providecommand \@href[1]{\@@startlink{#1}\@@href}%
\providecommand \@@href[1]{\endgroup#1\@@endlink}%
\providecommand \@sanitize@url [0]{\catcode `\\12\catcode `\$12\catcode `\&12\catcode `\#12\catcode `\^12\catcode `\_12\catcode `\%12\relax}%
\providecommand \@@startlink[1]{}%
\providecommand \@@endlink[0]{}%
\providecommand \url  [0]{\begingroup\@sanitize@url \@url }%
\providecommand \@url [1]{\endgroup\@href {#1}{\urlprefix }}%
\providecommand \urlprefix  [0]{URL }%
\providecommand \Eprint [0]{\href }%
\providecommand \doibase [0]{https://doi.org/}%
\providecommand \selectlanguage [0]{\@gobble}%
\providecommand \bibinfo  [0]{\@secondoftwo}%
\providecommand \bibfield  [0]{\@secondoftwo}%
\providecommand \translation [1]{[#1]}%
\providecommand \BibitemOpen [0]{}%
\providecommand \bibitemStop [0]{}%
\providecommand \bibitemNoStop [0]{.\EOS\space}%
\providecommand \EOS [0]{\spacefactor3000\relax}%
\providecommand \BibitemShut  [1]{\csname bibitem#1\endcsname}%
\let\auto@bib@innerbib\@empty
\bibitem [{\citenamefont {Tillmann}\ \emph {et~al.}(2013)\citenamefont {Tillmann}, \citenamefont {Daki{\'c}}, \citenamefont {Heilmann}, \citenamefont {Nolte}, \citenamefont {Szameit},\ and\ \citenamefont {Walther}}]{bosonsampling}%
  \BibitemOpen
  \bibfield  {author} {\bibinfo {author} {\bibfnamefont {M.}~\bibnamefont {Tillmann}}, \bibinfo {author} {\bibfnamefont {B.}~\bibnamefont {Daki{\'c}}}, \bibinfo {author} {\bibfnamefont {R.}~\bibnamefont {Heilmann}}, \bibinfo {author} {\bibfnamefont {S.}~\bibnamefont {Nolte}}, \bibinfo {author} {\bibfnamefont {A.}~\bibnamefont {Szameit}},\ and\ \bibinfo {author} {\bibfnamefont {P.}~\bibnamefont {Walther}},\ }\bibfield  {title} {\bibinfo {title} {Experimental boson sampling},\ }\href {https://doi.org/10.1038/nphoton.2013.102} {\bibfield  {journal} {\bibinfo  {journal} {Nature Photonics}\ }\textbf {\bibinfo {volume} {7}},\ \bibinfo {pages} {540} (\bibinfo {year} {2013})}\BibitemShut {NoStop}%
\bibitem [{\citenamefont {Knill}\ \emph {et~al.}(2001)\citenamefont {Knill}, \citenamefont {Laflamme},\ and\ \citenamefont {Milburn}}]{LOQC}%
  \BibitemOpen
  \bibfield  {author} {\bibinfo {author} {\bibfnamefont {E.}~\bibnamefont {Knill}}, \bibinfo {author} {\bibfnamefont {R.}~\bibnamefont {Laflamme}},\ and\ \bibinfo {author} {\bibfnamefont {G.~J.}\ \bibnamefont {Milburn}},\ }\bibfield  {title} {\bibinfo {title} {A scheme for efficient quantum computation with linear optics},\ }\href {https://doi.org/https://doi.org/10.1038/35051009} {\bibfield  {journal} {\bibinfo  {journal} {nature}\ }\textbf {\bibinfo {volume} {409}},\ \bibinfo {pages} {46} (\bibinfo {year} {2001})}\BibitemShut {NoStop}%
\bibitem [{\citenamefont {Bartolucci}\ \emph {et~al.}(2023)\citenamefont {Bartolucci}, \citenamefont {Birchall}, \citenamefont {Bomb{\'\i}n}, \citenamefont {Cable}, \citenamefont {Dawson}, \citenamefont {Gimeno-Segovia}, \citenamefont {Johnston}, \citenamefont {Kieling}, \citenamefont {Nickerson}, \citenamefont {Pant}, \citenamefont {Pastawski}, \citenamefont {Rudolph},\ and\ \citenamefont {Sparrow}}]{fusion}%
  \BibitemOpen
  \bibfield  {author} {\bibinfo {author} {\bibfnamefont {S.}~\bibnamefont {Bartolucci}}, \bibinfo {author} {\bibfnamefont {P.}~\bibnamefont {Birchall}}, \bibinfo {author} {\bibfnamefont {H.}~\bibnamefont {Bomb{\'\i}n}}, \bibinfo {author} {\bibfnamefont {H.}~\bibnamefont {Cable}}, \bibinfo {author} {\bibfnamefont {C.}~\bibnamefont {Dawson}}, \bibinfo {author} {\bibfnamefont {M.}~\bibnamefont {Gimeno-Segovia}}, \bibinfo {author} {\bibfnamefont {E.}~\bibnamefont {Johnston}}, \bibinfo {author} {\bibfnamefont {K.}~\bibnamefont {Kieling}}, \bibinfo {author} {\bibfnamefont {N.}~\bibnamefont {Nickerson}}, \bibinfo {author} {\bibfnamefont {M.}~\bibnamefont {Pant}}, \bibinfo {author} {\bibfnamefont {F.}~\bibnamefont {Pastawski}}, \bibinfo {author} {\bibfnamefont {T.}~\bibnamefont {Rudolph}},\ and\ \bibinfo {author} {\bibfnamefont {C.}~\bibnamefont {Sparrow}},\ }\bibfield  {title} {\bibinfo {title} {Fusion-based quantum computation},\ }\href {https://doi.org/10.1038/s41467-023-36493-1} {\bibfield  {journal} {\bibinfo
  {journal} {Nature Communications}\ }\textbf {\bibinfo {volume} {14}},\ \bibinfo {pages} {912} (\bibinfo {year} {2023})}\BibitemShut {NoStop}%
\bibitem [{\citenamefont {Motes}\ \emph {et~al.}(2015)\citenamefont {Motes}, \citenamefont {Olson}, \citenamefont {Rabeaux}, \citenamefont {Dowling}, \citenamefont {Olson},\ and\ \citenamefont {Rohde}}]{metrology}%
  \BibitemOpen
  \bibfield  {author} {\bibinfo {author} {\bibfnamefont {K.~R.}\ \bibnamefont {Motes}}, \bibinfo {author} {\bibfnamefont {J.~P.}\ \bibnamefont {Olson}}, \bibinfo {author} {\bibfnamefont {E.~J.}\ \bibnamefont {Rabeaux}}, \bibinfo {author} {\bibfnamefont {J.~P.}\ \bibnamefont {Dowling}}, \bibinfo {author} {\bibfnamefont {S.~J.}\ \bibnamefont {Olson}},\ and\ \bibinfo {author} {\bibfnamefont {P.~P.}\ \bibnamefont {Rohde}},\ }\bibfield  {title} {\bibinfo {title} {Linear optical quantum metrology with single photons: Exploiting spontaneously generated entanglement to beat the shot-noise limit},\ }\href {https://doi.org/10.1103/PhysRevLett.114.170802} {\bibfield  {journal} {\bibinfo  {journal} {Phys. Rev. Lett.}\ }\textbf {\bibinfo {volume} {114}},\ \bibinfo {pages} {170802} (\bibinfo {year} {2015})}\BibitemShut {NoStop}%
\bibitem [{\citenamefont {Fabre}\ and\ \citenamefont {Treps}(2020)}]{fabretreps}%
  \BibitemOpen
  \bibfield  {author} {\bibinfo {author} {\bibfnamefont {C.}~\bibnamefont {Fabre}}\ and\ \bibinfo {author} {\bibfnamefont {N.}~\bibnamefont {Treps}},\ }\bibfield  {title} {\bibinfo {title} {Modes and states in quantum optics},\ }\href {https://doi.org/10.1103/RevModPhys.92.035005} {\bibfield  {journal} {\bibinfo  {journal} {Rev. Mod. Phys.}\ }\textbf {\bibinfo {volume} {92}},\ \bibinfo {pages} {035005} (\bibinfo {year} {2020})}\BibitemShut {NoStop}%
\bibitem [{\citenamefont {Scully}\ and\ \citenamefont {Zubairy}(1997)}]{scully1997quantum}%
  \BibitemOpen
  \bibfield  {author} {\bibinfo {author} {\bibfnamefont {M.~O.}\ \bibnamefont {Scully}}\ and\ \bibinfo {author} {\bibfnamefont {M.~S.}\ \bibnamefont {Zubairy}},\ }\href@noop {} {\emph {\bibinfo {title} {Quantum optics}}}\ (\bibinfo  {publisher} {Cambridge university press},\ \bibinfo {year} {1997})\BibitemShut {NoStop}%
\bibitem [{\citenamefont {Niset}\ \emph {et~al.}(2009)\citenamefont {Niset}, \citenamefont {Fiur{\'a}{\v{s}}ek},\ and\ \citenamefont {Cerf}}]{niset2009no}%
  \BibitemOpen
  \bibfield  {author} {\bibinfo {author} {\bibfnamefont {J.}~\bibnamefont {Niset}}, \bibinfo {author} {\bibfnamefont {J.}~\bibnamefont {Fiur{\'a}{\v{s}}ek}},\ and\ \bibinfo {author} {\bibfnamefont {N.~J.}\ \bibnamefont {Cerf}},\ }\bibfield  {title} {\bibinfo {title} {No-go theorem for gaussian quantum error correction},\ }\href {https://doi.org/https://doi.org/10.1103/PhysRevLett.102.120501} {\bibfield  {journal} {\bibinfo  {journal} {Physical review letters}\ }\textbf {\bibinfo {volume} {102}},\ \bibinfo {pages} {120501} (\bibinfo {year} {2009})}\BibitemShut {NoStop}%
\bibitem [{\citenamefont {Petz}(1986)}]{petz1986sufficient}%
  \BibitemOpen
  \bibfield  {author} {\bibinfo {author} {\bibfnamefont {D.}~\bibnamefont {Petz}},\ }\bibfield  {title} {\bibinfo {title} {Sufficient subalgebras and the relative entropy of states of a von neumann algebra},\ }\href {https://doi.org/https://doi.org/10.1007/BF01212345} {\bibfield  {journal} {\bibinfo  {journal} {Communications in mathematical physics}\ }\textbf {\bibinfo {volume} {105}},\ \bibinfo {pages} {123} (\bibinfo {year} {1986})}\BibitemShut {NoStop}%
\bibitem [{\citenamefont {Petz}(1988)}]{petz1988sufficiency}%
  \BibitemOpen
  \bibfield  {author} {\bibinfo {author} {\bibfnamefont {D.}~\bibnamefont {Petz}},\ }\bibfield  {title} {\bibinfo {title} {Sufficiency of channels over von neumann algebras},\ }\href {https://doi.org/https://doi.org/10.1093/qmath/39.1.97} {\bibfield  {journal} {\bibinfo  {journal} {The Quarterly Journal of Mathematics}\ }\textbf {\bibinfo {volume} {39}},\ \bibinfo {pages} {97} (\bibinfo {year} {1988})}\BibitemShut {NoStop}%
\bibitem [{\citenamefont {Barnum}\ and\ \citenamefont {Knill}(2002)}]{nearoptimal1}%
  \BibitemOpen
  \bibfield  {author} {\bibinfo {author} {\bibfnamefont {H.}~\bibnamefont {Barnum}}\ and\ \bibinfo {author} {\bibfnamefont {E.}~\bibnamefont {Knill}},\ }\bibfield  {title} {\bibinfo {title} {Reversing quantum dynamics with near-optimal quantum and classical fidelity},\ }\href {https://doi.org/https://doi.org/10.1063/1.1459754} {\bibfield  {journal} {\bibinfo  {journal} {Journal of Mathematical Physics}\ }\textbf {\bibinfo {volume} {43}},\ \bibinfo {pages} {2097} (\bibinfo {year} {2002})}\BibitemShut {NoStop}%
\bibitem [{\citenamefont {Ng}\ and\ \citenamefont {Mandayam}(2010)}]{nearpotimal2}%
  \BibitemOpen
  \bibfield  {author} {\bibinfo {author} {\bibfnamefont {H.~K.}\ \bibnamefont {Ng}}\ and\ \bibinfo {author} {\bibfnamefont {P.}~\bibnamefont {Mandayam}},\ }\bibfield  {title} {\bibinfo {title} {Simple approach to approximate quantum error correction based on the transpose channel},\ }\href {https://doi.org/https://doi.org/10.1103/PhysRevA.81.062342} {\bibfield  {journal} {\bibinfo  {journal} {Physical Review A—Atomic, Molecular, and Optical Physics}\ }\textbf {\bibinfo {volume} {81}},\ \bibinfo {pages} {062342} (\bibinfo {year} {2010})}\BibitemShut {NoStop}%
\bibitem [{\citenamefont {Zheng}\ \emph {et~al.}(2024)\citenamefont {Zheng}, \citenamefont {He}, \citenamefont {Lee},\ and\ \citenamefont {Jiang}}]{nearoptimal3}%
  \BibitemOpen
  \bibfield  {author} {\bibinfo {author} {\bibfnamefont {G.}~\bibnamefont {Zheng}}, \bibinfo {author} {\bibfnamefont {W.}~\bibnamefont {He}}, \bibinfo {author} {\bibfnamefont {G.}~\bibnamefont {Lee}},\ and\ \bibinfo {author} {\bibfnamefont {L.}~\bibnamefont {Jiang}},\ }\bibfield  {title} {\bibinfo {title} {Near-optimal performance of quantum error correction codes},\ }\href {https://doi.org/https://doi.org/10.1103/PhysRevLett.132.250602} {\bibfield  {journal} {\bibinfo  {journal} {Physical Review Letters}\ }\textbf {\bibinfo {volume} {132}},\ \bibinfo {pages} {250602} (\bibinfo {year} {2024})}\BibitemShut {NoStop}%
\bibitem [{\citenamefont {Li}\ \emph {et~al.}(2025{\natexlab{a}})\citenamefont {Li}, \citenamefont {Wang}, \citenamefont {Zheng}, \citenamefont {Wong},\ and\ \citenamefont {Jiang}}]{nearoptimal4}%
  \BibitemOpen
  \bibfield  {author} {\bibinfo {author} {\bibfnamefont {B.}~\bibnamefont {Li}}, \bibinfo {author} {\bibfnamefont {Z.}~\bibnamefont {Wang}}, \bibinfo {author} {\bibfnamefont {G.}~\bibnamefont {Zheng}}, \bibinfo {author} {\bibfnamefont {Y.}~\bibnamefont {Wong}},\ and\ \bibinfo {author} {\bibfnamefont {L.}~\bibnamefont {Jiang}},\ }\bibfield  {title} {\bibinfo {title} {Optimality condition for the petz map},\ }\href {https://doi.org/https://doi.org/10.1103/PhysRevLett.134.200602} {\bibfield  {journal} {\bibinfo  {journal} {Physical Review Letters}\ }\textbf {\bibinfo {volume} {134}},\ \bibinfo {pages} {200602} (\bibinfo {year} {2025}{\natexlab{a}})}\BibitemShut {NoStop}%
\bibitem [{\citenamefont {Biswas}\ \emph {et~al.}(2024)\citenamefont {Biswas}, \citenamefont {Vaidya},\ and\ \citenamefont {Mandayam}}]{Petzerrorcorrection}%
  \BibitemOpen
  \bibfield  {author} {\bibinfo {author} {\bibfnamefont {D.}~\bibnamefont {Biswas}}, \bibinfo {author} {\bibfnamefont {G.~M.}\ \bibnamefont {Vaidya}},\ and\ \bibinfo {author} {\bibfnamefont {P.}~\bibnamefont {Mandayam}},\ }\bibfield  {title} {\bibinfo {title} {Noise-adapted recovery circuits for quantum error correction},\ }\href {https://doi.org/https://doi.org/10.1103/PhysRevResearch.6.043034} {\bibfield  {journal} {\bibinfo  {journal} {Physical Review Research}\ }\textbf {\bibinfo {volume} {6}},\ \bibinfo {pages} {043034} (\bibinfo {year} {2024})}\BibitemShut {NoStop}%
\bibitem [{\citenamefont {Beigi}\ \emph {et~al.}(2016)\citenamefont {Beigi}, \citenamefont {Datta},\ and\ \citenamefont {Leditzky}}]{dataprocessing}%
  \BibitemOpen
  \bibfield  {author} {\bibinfo {author} {\bibfnamefont {S.}~\bibnamefont {Beigi}}, \bibinfo {author} {\bibfnamefont {N.}~\bibnamefont {Datta}},\ and\ \bibinfo {author} {\bibfnamefont {F.}~\bibnamefont {Leditzky}},\ }\bibfield  {title} {\bibinfo {title} {Decoding quantum information via the petz recovery map},\ }\bibfield  {journal} {\bibinfo  {journal} {Journal of Mathematical Physics}\ }\textbf {\bibinfo {volume} {57}},\ \href {https://doi.org/https://doi.org/10.1063/1.4961515} {https://doi.org/10.1063/1.4961515} (\bibinfo {year} {2016})\BibitemShut {NoStop}%
\bibitem [{\citenamefont {Kwon}\ and\ \citenamefont {Kim}(2019)}]{kwon-kim}%
  \BibitemOpen
  \bibfield  {author} {\bibinfo {author} {\bibfnamefont {H.}~\bibnamefont {Kwon}}\ and\ \bibinfo {author} {\bibfnamefont {M.~S.}\ \bibnamefont {Kim}},\ }\bibfield  {title} {\bibinfo {title} {Fluctuation theorems for a quantum channel},\ }\href {https://doi.org/10.1103/PhysRevX.9.031029} {\bibfield  {journal} {\bibinfo  {journal} {Phys. Rev. X}\ }\textbf {\bibinfo {volume} {9}},\ \bibinfo {pages} {031029} (\bibinfo {year} {2019})}\BibitemShut {NoStop}%
\bibitem [{\citenamefont {Aw}\ \emph {et~al.}(2021)\citenamefont {Aw}, \citenamefont {Buscemi},\ and\ \citenamefont {Scarani}}]{fluctuationtheorem}%
  \BibitemOpen
  \bibfield  {author} {\bibinfo {author} {\bibfnamefont {C.~C.}\ \bibnamefont {Aw}}, \bibinfo {author} {\bibfnamefont {F.}~\bibnamefont {Buscemi}},\ and\ \bibinfo {author} {\bibfnamefont {V.}~\bibnamefont {Scarani}},\ }\bibfield  {title} {\bibinfo {title} {Fluctuation theorems with retrodiction rather than reverse processes},\ }\href {https://doi.org/10.1116/5.0060893} {\bibfield  {journal} {\bibinfo  {journal} {AVS Quantum Science}\ }\textbf {\bibinfo {volume} {3}},\ \bibinfo {pages} {045601} (\bibinfo {year} {2021})},\ \Eprint {https://arxiv.org/abs/https://pubs.aip.org/avs/aqs/article-pdf/doi/10.1116/5.0060893/19738704/045601\_1\_online.pdf} {https://pubs.aip.org/avs/aqs/article-pdf/doi/10.1116/5.0060893/19738704/045601\_1\_online.pdf} \BibitemShut {NoStop}%
\bibitem [{\citenamefont {Buscemi}\ \emph {et~al.}(2023)\citenamefont {Buscemi}, \citenamefont {Schindler},\ and\ \citenamefont {{\v{S}}afr{\'a}nek}}]{entropy}%
  \BibitemOpen
  \bibfield  {author} {\bibinfo {author} {\bibfnamefont {F.}~\bibnamefont {Buscemi}}, \bibinfo {author} {\bibfnamefont {J.}~\bibnamefont {Schindler}},\ and\ \bibinfo {author} {\bibfnamefont {D.}~\bibnamefont {{\v{S}}afr{\'a}nek}},\ }\bibfield  {title} {\bibinfo {title} {Observational entropy, coarse-grained states, and the petz recovery map: information-theoretic properties and bounds},\ }\href {https://doi.org/10.1088/1367-2630/accd11} {\bibfield  {journal} {\bibinfo  {journal} {New Journal of Physics}\ }\textbf {\bibinfo {volume} {25}},\ \bibinfo {pages} {053002} (\bibinfo {year} {2023})}\BibitemShut {NoStop}%
\bibitem [{\citenamefont {Li}\ \emph {et~al.}(2025{\natexlab{b}})\citenamefont {Li}, \citenamefont {Xie}, \citenamefont {Kwon}, \citenamefont {Zhao}, \citenamefont {Kim},\ and\ \citenamefont {Zhang}}]{kwon-zhang}%
  \BibitemOpen
  \bibfield  {author} {\bibinfo {author} {\bibfnamefont {H.}~\bibnamefont {Li}}, \bibinfo {author} {\bibfnamefont {J.}~\bibnamefont {Xie}}, \bibinfo {author} {\bibfnamefont {H.}~\bibnamefont {Kwon}}, \bibinfo {author} {\bibfnamefont {Y.}~\bibnamefont {Zhao}}, \bibinfo {author} {\bibfnamefont {M.~S.}\ \bibnamefont {Kim}},\ and\ \bibinfo {author} {\bibfnamefont {L.}~\bibnamefont {Zhang}},\ }\bibfield  {title} {\bibinfo {title} {Experimental demonstration of generalized quantum fluctuation theorems in the presence of coherence},\ }\href {https://doi.org/10.1126/sciadv.adq6014} {\bibfield  {journal} {\bibinfo  {journal} {Science Advances}\ }\textbf {\bibinfo {volume} {11}},\ \bibinfo {pages} {eadq6014} (\bibinfo {year} {2025}{\natexlab{b}})},\ \Eprint {https://arxiv.org/abs/https://www.science.org/doi/pdf/10.1126/sciadv.adq6014} {https://www.science.org/doi/pdf/10.1126/sciadv.adq6014} \BibitemShut {NoStop}%
\bibitem [{\citenamefont {Png}\ and\ \citenamefont {Scarani}(2025)}]{wenhan}%
  \BibitemOpen
  \bibfield  {author} {\bibinfo {author} {\bibfnamefont {W.-H.}\ \bibnamefont {Png}}\ and\ \bibinfo {author} {\bibfnamefont {V.}~\bibnamefont {Scarani}},\ }\bibfield  {title} {\bibinfo {title} {Petz recovery maps of single-qubit decoherence channels in an ion trap quantum processor},\ }\href {https://doi.org/10.1103/7f8x-n2np} {\bibfield  {journal} {\bibinfo  {journal} {Phys. Rev. A}\ }\textbf {\bibinfo {volume} {112}},\ \bibinfo {pages} {022613} (\bibinfo {year} {2025})}\BibitemShut {NoStop}%
\bibitem [{\citenamefont {Singh}\ \emph {et~al.}(2025)\citenamefont {Singh}, \citenamefont {Sahani}, \citenamefont {Jagadish}, \citenamefont {Lautenbacher}, \citenamefont {Bernardes},\ and\ \citenamefont {Dorai}}]{leaNMR}%
  \BibitemOpen
  \bibfield  {author} {\bibinfo {author} {\bibfnamefont {G.}~\bibnamefont {Singh}}, \bibinfo {author} {\bibfnamefont {R.}~\bibnamefont {Sahani}}, \bibinfo {author} {\bibfnamefont {V.}~\bibnamefont {Jagadish}}, \bibinfo {author} {\bibfnamefont {L.}~\bibnamefont {Lautenbacher}}, \bibinfo {author} {\bibfnamefont {N.}~\bibnamefont {Bernardes}},\ and\ \bibinfo {author} {\bibfnamefont {K.}~\bibnamefont {Dorai}},\ }\href@noop {} {\bibinfo {title} {Realizing the petz recovery map on an nmr quantum processor}} (\bibinfo {year} {2025}),\ \Eprint {https://arxiv.org/abs/arxiv:2508.08998} {arXiv:arxiv:2508.08998} \BibitemShut {NoStop}%
\bibitem [{\citenamefont {Lami}\ \emph {et~al.}(2018)\citenamefont {Lami}, \citenamefont {Das},\ and\ \citenamefont {Wilde}}]{GaussianPetz2}%
  \BibitemOpen
  \bibfield  {author} {\bibinfo {author} {\bibfnamefont {L.}~\bibnamefont {Lami}}, \bibinfo {author} {\bibfnamefont {S.}~\bibnamefont {Das}},\ and\ \bibinfo {author} {\bibfnamefont {M.~M.}\ \bibnamefont {Wilde}},\ }\bibfield  {title} {\bibinfo {title} {Approximate reversal of quantum gaussian dynamics},\ }\href {https://doi.org/10.1088/1751-8121/aaad26} {\bibfield  {journal} {\bibinfo  {journal} {Journal of Physics A: Mathematical and Theoretical}\ }\textbf {\bibinfo {volume} {51}},\ \bibinfo {pages} {125301} (\bibinfo {year} {2018})}\BibitemShut {NoStop}%
\bibitem [{\citenamefont {Weedbrook}\ \emph {et~al.}(2012{\natexlab{a}})\citenamefont {Weedbrook}, \citenamefont {Pirandola}, \citenamefont {García-Patrón}, \citenamefont {Cerf}, \citenamefont {Ralph}, \citenamefont {Shapiro},\ and\ \citenamefont {Lloyd}}]{Gaussianbook}%
  \BibitemOpen
  \bibfield  {author} {\bibinfo {author} {\bibfnamefont {C.}~\bibnamefont {Weedbrook}}, \bibinfo {author} {\bibfnamefont {S.}~\bibnamefont {Pirandola}}, \bibinfo {author} {\bibfnamefont {R.}~\bibnamefont {García-Patrón}}, \bibinfo {author} {\bibfnamefont {N.~J.}\ \bibnamefont {Cerf}}, \bibinfo {author} {\bibfnamefont {T.~C.}\ \bibnamefont {Ralph}}, \bibinfo {author} {\bibfnamefont {J.~H.}\ \bibnamefont {Shapiro}},\ and\ \bibinfo {author} {\bibfnamefont {S.}~\bibnamefont {Lloyd}},\ }\bibfield  {title} {\bibinfo {title} {Gaussian quantum information},\ }\href {https://doi.org/10.1103/revmodphys.84.621} {\bibfield  {journal} {\bibinfo  {journal} {Reviews of Modern Physics}\ }\textbf {\bibinfo {volume} {84}},\ \bibinfo {pages} {621–669} (\bibinfo {year} {2012}{\natexlab{a}})}\BibitemShut {NoStop}%
\bibitem [{\citenamefont {Robertson}(1929)}]{robertson1929uncertainty}%
  \BibitemOpen
  \bibfield  {author} {\bibinfo {author} {\bibfnamefont {H.~P.}\ \bibnamefont {Robertson}},\ }\bibfield  {title} {\bibinfo {title} {The uncertainty principle},\ }\href {https://doi.org/https://doi.org/10.1103/PhysRev.34.163} {\bibfield  {journal} {\bibinfo  {journal} {Physical Review}\ }\textbf {\bibinfo {volume} {34}},\ \bibinfo {pages} {163} (\bibinfo {year} {1929})}\BibitemShut {NoStop}%
\bibitem [{\citenamefont {Weedbrook}\ \emph {et~al.}(2012{\natexlab{b}})\citenamefont {Weedbrook}, \citenamefont {Pirandola}, \citenamefont {Garc{\'\i}a-Patr{\'o}n}, \citenamefont {Cerf}, \citenamefont {Ralph}, \citenamefont {Shapiro},\ and\ \citenamefont {Lloyd}}]{Gaussianinformation}%
  \BibitemOpen
  \bibfield  {author} {\bibinfo {author} {\bibfnamefont {C.}~\bibnamefont {Weedbrook}}, \bibinfo {author} {\bibfnamefont {S.}~\bibnamefont {Pirandola}}, \bibinfo {author} {\bibfnamefont {R.}~\bibnamefont {Garc{\'\i}a-Patr{\'o}n}}, \bibinfo {author} {\bibfnamefont {N.~J.}\ \bibnamefont {Cerf}}, \bibinfo {author} {\bibfnamefont {T.~C.}\ \bibnamefont {Ralph}}, \bibinfo {author} {\bibfnamefont {J.~H.}\ \bibnamefont {Shapiro}},\ and\ \bibinfo {author} {\bibfnamefont {S.}~\bibnamefont {Lloyd}},\ }\bibfield  {title} {\bibinfo {title} {Gaussian quantum information},\ }\href {https://doi.org/https://doi.org/10.1103/RevModPhys.84.621} {\bibfield  {journal} {\bibinfo  {journal} {Reviews of Modern Physics}\ }\textbf {\bibinfo {volume} {84}},\ \bibinfo {pages} {621} (\bibinfo {year} {2012}{\natexlab{b}})}\BibitemShut {NoStop}%
\bibitem [{\citenamefont {Serafini}(2023)}]{serafini2023qcv}%
  \BibitemOpen
  \bibfield  {author} {\bibinfo {author} {\bibfnamefont {A.}~\bibnamefont {Serafini}},\ }\href {https://doi.org/https://doi.org/10.1201/9781003250975} {\emph {\bibinfo {title} {Quantum continuous variables: a primer of theoretical methods}}}\ (\bibinfo  {publisher} {CRC press},\ \bibinfo {year} {2023})\BibitemShut {NoStop}%
\bibitem [{\citenamefont {Parzygnat}\ and\ \citenamefont {Buscemi}(2023)}]{Parzygnat2023axiomsretrodiction}%
  \BibitemOpen
  \bibfield  {author} {\bibinfo {author} {\bibfnamefont {A.~J.}\ \bibnamefont {Parzygnat}}\ and\ \bibinfo {author} {\bibfnamefont {F.}~\bibnamefont {Buscemi}},\ }\bibfield  {title} {\bibinfo {title} {Axioms for retrodiction: achieving time-reversal symmetry with a prior},\ }\href {https://doi.org/10.22331/q-2023-05-23-1013} {\bibfield  {journal} {\bibinfo  {journal} {{Quantum}}\ }\textbf {\bibinfo {volume} {7}},\ \bibinfo {pages} {1013} (\bibinfo {year} {2023})}\BibitemShut {NoStop}%
\bibitem [{\citenamefont {Bai}\ \emph {et~al.}(2025)\citenamefont {Bai}, \citenamefont {Buscemi},\ and\ \citenamefont {Scarani}}]{minchange}%
  \BibitemOpen
  \bibfield  {author} {\bibinfo {author} {\bibfnamefont {G.}~\bibnamefont {Bai}}, \bibinfo {author} {\bibfnamefont {F.}~\bibnamefont {Buscemi}},\ and\ \bibinfo {author} {\bibfnamefont {V.}~\bibnamefont {Scarani}},\ }\bibfield  {title} {\bibinfo {title} {Quantum bayes' rule and petz transpose map from the minimum change principle},\ }\href {https://doi.org/10.1103/5n4p-bxhm} {\bibfield  {journal} {\bibinfo  {journal} {Phys. Rev. Lett.}\ }\textbf {\bibinfo {volume} {135}},\ \bibinfo {pages} {090203} (\bibinfo {year} {2025})}\BibitemShut {NoStop}%
\bibitem [{\citenamefont {Hayden}\ \emph {et~al.}(2004)\citenamefont {Hayden}, \citenamefont {Jozsa}, \citenamefont {Petz},\ and\ \citenamefont {Winter}}]{hayden2004qmarkov}%
  \BibitemOpen
  \bibfield  {author} {\bibinfo {author} {\bibfnamefont {P.}~\bibnamefont {Hayden}}, \bibinfo {author} {\bibfnamefont {R.}~\bibnamefont {Jozsa}}, \bibinfo {author} {\bibfnamefont {D.}~\bibnamefont {Petz}},\ and\ \bibinfo {author} {\bibfnamefont {A.}~\bibnamefont {Winter}},\ }\bibfield  {title} {\bibinfo {title} {Structure of states which satisfy strong subadditivity of quantum entropy with equality},\ }\href {https://doi.org/10.1007/s00220-004-1049-z} {\bibfield  {journal} {\bibinfo  {journal} {Communications in mathematical physics}\ }\textbf {\bibinfo {volume} {246}},\ \bibinfo {pages} {359} (\bibinfo {year} {2004})}\BibitemShut {NoStop}%
\bibitem [{\citenamefont {Wilde}(2015)}]{recoverability}%
  \BibitemOpen
  \bibfield  {author} {\bibinfo {author} {\bibfnamefont {M.~M.}\ \bibnamefont {Wilde}},\ }\bibfield  {title} {\bibinfo {title} {Recoverability in quantum information theory},\ }\href {https://doi.org/https://doi.org/10.1098/rspa.2015.0338} {\bibfield  {journal} {\bibinfo  {journal} {Proceedings of the Royal Society A: Mathematical, Physical and Engineering Sciences}\ }\textbf {\bibinfo {volume} {471}},\ \bibinfo {pages} {20150338} (\bibinfo {year} {2015})}\BibitemShut {NoStop}%
\bibitem [{\citenamefont {Uhlmann}(1976)}]{uhlmann1976fidelity}%
  \BibitemOpen
  \bibfield  {author} {\bibinfo {author} {\bibfnamefont {A.}~\bibnamefont {Uhlmann}},\ }\bibfield  {title} {\bibinfo {title} {The ``transition probability'' in the state space of a*-algebra},\ }\href {https://doi.org/https://doi.org/10.1016/0034-4877(76)90060-4} {\bibfield  {journal} {\bibinfo  {journal} {Reports on Mathematical Physics}\ }\textbf {\bibinfo {volume} {9}},\ \bibinfo {pages} {273} (\bibinfo {year} {1976})}\BibitemShut {NoStop}%
\bibitem [{\citenamefont {Jozsa}(1994)}]{jozsa1994fidelity}%
  \BibitemOpen
  \bibfield  {author} {\bibinfo {author} {\bibfnamefont {R.}~\bibnamefont {Jozsa}},\ }\bibfield  {title} {\bibinfo {title} {Fidelity for mixed quantum states},\ }\href {https://doi.org/https://doi.org/10.1080/09500349414552171} {\bibfield  {journal} {\bibinfo  {journal} {Journal of modern optics}\ }\textbf {\bibinfo {volume} {41}},\ \bibinfo {pages} {2315} (\bibinfo {year} {1994})}\BibitemShut {NoStop}%
\bibitem [{\citenamefont {Vidal}\ and\ \citenamefont {Werner}(2002)}]{vidal2002computable}%
  \BibitemOpen
  \bibfield  {author} {\bibinfo {author} {\bibfnamefont {G.}~\bibnamefont {Vidal}}\ and\ \bibinfo {author} {\bibfnamefont {R.~F.}\ \bibnamefont {Werner}},\ }\bibfield  {title} {\bibinfo {title} {Computable measure of entanglement},\ }\href {https://doi.org/https://doi.org/10.1103/PhysRevA.65.032314} {\bibfield  {journal} {\bibinfo  {journal} {Physical Review A}\ }\textbf {\bibinfo {volume} {65}},\ \bibinfo {pages} {032314} (\bibinfo {year} {2002})}\BibitemShut {NoStop}%
\bibitem [{\citenamefont {Adesso}\ \emph {et~al.}(2004)\citenamefont {Adesso}, \citenamefont {Serafini},\ and\ \citenamefont {Illuminati}}]{adesso2004extremal}%
  \BibitemOpen
  \bibfield  {author} {\bibinfo {author} {\bibfnamefont {G.}~\bibnamefont {Adesso}}, \bibinfo {author} {\bibfnamefont {A.}~\bibnamefont {Serafini}},\ and\ \bibinfo {author} {\bibfnamefont {F.}~\bibnamefont {Illuminati}},\ }\bibfield  {title} {\bibinfo {title} {Extremal entanglement and mixedness in continuous variable systems},\ }\href {https://doi.org/https://doi.org/10.1103/PhysRevA.70.022318} {\bibfield  {journal} {\bibinfo  {journal} {Physical Review A—Atomic, Molecular, and Optical Physics}\ }\textbf {\bibinfo {volume} {70}},\ \bibinfo {pages} {022318} (\bibinfo {year} {2004})}\BibitemShut {NoStop}%
\bibitem [{\citenamefont {Marian}\ and\ \citenamefont {Marian}(2012)}]{marian2012fidelity_Gaussian}%
  \BibitemOpen
  \bibfield  {author} {\bibinfo {author} {\bibfnamefont {P.}~\bibnamefont {Marian}}\ and\ \bibinfo {author} {\bibfnamefont {T.~A.}\ \bibnamefont {Marian}},\ }\bibfield  {title} {\bibinfo {title} {Uhlmann fidelity between two-mode gaussian states},\ }\href {https://doi.org/https://doi.org/10.1103/PhysRevA.86.022340} {\bibfield  {journal} {\bibinfo  {journal} {Physical Review A—Atomic, Molecular, and Optical Physics}\ }\textbf {\bibinfo {volume} {86}},\ \bibinfo {pages} {022340} (\bibinfo {year} {2012})}\BibitemShut {NoStop}%
\end{thebibliography}%

\appendix
\widetext
\setcounter{rst}{0}

\section{A General solution to Gaussian lossy channel}
\label{general solution}
In this part, we focus on a general Gaussian lossy channel, which can be expressed as a beam splitter with transmissivity $\eta$, a Gaussian environment $\xi$ with covariance matrix $V_\xi$ and mean vector $\mathbf{\overline{\mathbf{r}}}_\xi$. The environment will later be traced out. Thus the Gaussian channel $\mathcal{N}$ is can be expressed with
\begin{equation}
    \label{eq:general lossy model}
    X_\mathcal{N} = \sqrt{\eta}I\,,\; Y_\mathcal{N} = (1-\eta)V_{\xi}\,,\; \mathbf{d}_\mathcal{N} = \sqrt{1-\eta}\,\overline{\mathbf{r}}_{\xi}.
\end{equation}
If the prior $\sigma$ is also taken as Gaussian, the Petz recovery map is a Gaussian channel given by \cref{eq:petz_XY}. 

We focus on nontrivial lossy channels $0<\eta<1$. Plugging Eq.~\eqref{eq:general lossy model} into Eq.~\eqref{eq:petz_XY} and using $M\Omega M^T= (\det M)\Omega$ for single mode Gaussian state, one finds first
\begin{align}
    X_{\mathcal{P}} &= 
    \left(\eta \,
    \frac{1-(\det V_{\sigma})^{-1}}{1-(\det V_{\mathcal{N}(\sigma)})^{-1}}
    \right)^{\frac{1}{2}}
    V_\sigma V_{\mathcal{N}(\sigma)}^{-1}.
    \label{eq:Petz_general}
\end{align}

The Petz recovery map is another beam splitter under the conditions such that:
\begin{align}
         &V_{\sigma}\propto V_{\mathcal{N}(\sigma)}\; \textrm{[equivalently, $V_\sigma \propto V_\xi$]}\,, \qand \label{eq:cond_app1}\\
        &\max\left(1,\sqrt{\det V_
        \xi}-\sqrt{\frac{\det V_\xi -1}{\eta}} \right)\le \sqrt{\det V_\sigma} \le \sqrt{\det V_
        \xi}+\sqrt{\frac{\det V_\xi -1}{\eta}}\,. \label{eq:cond_app2}
\end{align}

Specifically, under the conditions \eqref{eq:cond_app1} and \eqref{eq:cond_app2}, the Petz recovery map $\mathcal{P}_{\mathcal{N},\sigma}$ can be realized by a beam splitter with transmissivity $\eta'$ and a Gaussian ancilla $\xi'$, characterized by
\begin{align}
    X_{\mathcal{P}} = \sqrt{\eta'} I, 
    Y_\mathcal{P} = (1-\eta')V_{\xi'},
    \mathbf{d}_{\mathcal{P}} = \sqrt{1-\eta'} \overline{\mathbf{r}}_{\xi'},
\end{align}
where $\eta'=\eta'(\sigma)$ is given by 
\begin{align}
    \eta' = \eta\,\frac{\det V_\sigma-1}{\det V_{\mathcal{N}(\sigma)}-1},
\end{align}
and $\xi'=\xi'(\sigma)$ is a Gaussian state $(\overline{\mathbf{r}}_{\xi'},V_{\xi'})$ such that 
\begin{align}
\begin{aligned}
    \overline{\mathbf{r}}_{\xi'} &= \frac{1}{\sqrt{1-\eta'}}\left(\overline{\mathbf{r}}_{\sigma}-\sqrt{\eta'}(\sqrt{\eta} \overline{\mathbf{r}}_{\sigma}+\sqrt{1-\eta} \overline{\mathbf{r}}_{\xi})\right),\\
    V_{\xi'} &= \frac{1}{1-\eta'}\,\left(1-\eta'\sqrt{\frac{\det V_{\mathcal{N}(\sigma)}}{\det V_{\sigma}}}\right)\,V_\sigma.
\end{aligned}
\end{align}

Denoting $x:= \sqrt{\det V_\sigma}, y:=\sqrt{\det V_{\mathcal{N}(\sigma)}}$, we can express $\det V_{\xi'(\sigma)}$ in terms of $x$ and $y$:
    \begin{align}
        \det V_{\xi'(\sigma)} &= \left(1-\eta\left(\frac{x^2-1}{y^2-1}\right)\right)^{-2}\left(x-\eta\left(\frac{x^2-1}{y^2-1}\right)y\right)^2\\
        &= \left(y^2-1-\eta\left(x^2-1\right)\right)^{-2}\left(x(y^2-1)-\eta y(x^2-1)\right)^2
    \end{align}
    Then, it is readily seen that $\det V_{\xi'(\sigma)}\ge 1$:
    \begin{align}
        \det V_{\xi'(\sigma)}\ge1 &\Leftrightarrow \left(x(y^2-1)-\eta y(x^2-1)\right)^2 \ge \left(y^2-1-\eta\left(x^2-1\right)\right)^{2}\\
        &\Leftrightarrow \left( (x-1)(y^2-1)-\eta(x^2-1)(y-1)\right)\left( (x+1)(y^2-1)-\eta(x^2-1)(y+1)\right)\ge 0\\
        &\Leftrightarrow (x^2-1)(y^2-1)\left( y+1 - \eta(x+1) \right)\left( y-1 - \eta(x-1) \right) \ge 0\\
        &\Leftrightarrow (x^2-1)(y^2-1)\left( (1-\eta)b + 1-\eta \right)\left( (1-\eta)b - (1-\eta) \right) \ge 0 \quad (\because y=\eta x + (1-\eta)b)\\
        &\Leftrightarrow (x^2-1)(y^2-1)(1-\eta)^2(b^2-1) \ge 0,
    \end{align}
    where we used $A^2-B^2=(A-B)(A+B)$ in the second line. Hence, we conclude that $\det V_{\xi'(\sigma)}\ge 1$ using the fact that $\sigma,\xi,\mathcal{N}(\sigma)$ are physical states, i.e., $\det V_\sigma \geq 1, \det V_\xi \geq 1, \det V_{\mathcal{N}(\sigma)} \geq 1$. 

    In particular, when $\xi$ and $\sigma$ are both thermal states, we can reproduce what is presented in \cref{result1}. 

\section{Comparison between Petz recovery maps and the trivial recovery protocols $\mathcal{R}^{(0)}, \mathcal{R}^{(1)}_\sigma$} \label{app:fidelity}
In this part, we will often use $\mathcal{P}$ to denote $\mathcal{P}_{\mathcal{N},\sigma}$ for the sake of brevity. The forward channel $\mathcal{N}$ and its Petz recovery map $\mathcal{P}_{\mathcal{N},\sigma}$ are characterized by 
\begin{align}
    X_\mathcal{N} = \sqrt{\eta}I, Y_\mathcal{N} = (1-\eta)V_\xi, \,\text{and} \,\, \mathbf{d}_\mathcal{N} = 0\\
    X_{\mathcal{P}} = \sqrt{\eta'} I, 
    Y_\mathcal{P} = V_\sigma -\eta' V_{\mathcal{N}(\sigma)}, \,\text{and} \,\,
    \mathbf{d}_{\mathcal{P}} = 0. \label{eq:Petz_eta'}
\end{align}

The fidelity $F(\rho_1,\rho_2)$ between $\rho_1$ and $\rho_2$ is defined as $F(\rho_1,\rho_2)\equiv \left(\Tr\sqrt{\sqrt{\rho_2}\rho_1\sqrt{\rho_2}}\right)^2$~\cite{uhlmann1976fidelity,jozsa1994fidelity}. Following Ref.~\cite{marian2012fidelity_Gaussian}, the fidelity between two single-mode Gaussian states $\rho_1,\rho_2$ can be computed via a closed form expression 
    \begin{align}
        F(\rho_1,\rho_2) = \frac{1}{\Gamma(V_1,V_2)}e^{-\frac{1}{4}(\overline{\mathbf{r}}_2-\overline{\mathbf{r}}_1)^T(V_1+V_2)^T (\overline{\mathbf{r}}_2-\overline{\mathbf{r}}_1)},
    \end{align}
    in terms of mean vectors $\overline{\mathbf{r}}_1,\overline{\mathbf{r}}_2$ and covariance matrices $V_1,V_2$ of the Gaussian states $\rho_1,\rho_2$, respectively, with
    \begin{align}
        \Gamma(V_1,V_2) &\equiv  \sqrt{\Delta(V_1,V_2)+\Lambda(V_1,V_2)}-\sqrt{\Lambda(V_1,V_2)}, \label{eq:Gamma}\\
        \Delta(V_1,V_2) &\equiv \frac{1}{4}\det (V_1+V_2), \label{eq:Delta}\\
        \Lambda(V_1,V_2) &\equiv \frac{1}{4} \det(V_1+i\Omega)\det(V_2+i\Omega) \nonumber\\
            &=\frac{1}{4} (\det V_1-1)(\det V_2-1) \label{eq:Lambda}. 
    \end{align}

\subsection{The recovery protocol $\mathcal{R}^{(0)}$}\label{app:fidelity0}

\setcounter{rst}{1}
\begin{rst}
    Let $\rho$ and $\sigma$ be thermal states. Then, 
    \begin{align}
        F(\rho, \mathcal{R}^{(0)}\circ\mathcal{N}(\rho))= F(\rho, \mathcal{N}(\rho)) \le F(\rho, \mathcal{P}_{\mathcal{N},\sigma}\circ \mathcal{N}(\rho)), 
    \end{align}
    when $\max(1, \min (z_0,z_1)) \le g(\sigma) \leq \max (z_0, z_1)$, defining 
    \begin{align}
        f(z)&:= \frac{1}{2}\left( z\sqrt{\det V_\rho}+1 - (\det V_\rho-1)^\frac{1}{2}(z^2-1)^\frac{1}{2} \right)\\
        g(\sigma)&:= \eta\eta'(\sigma)\sqrt{\det V_\rho} + \left(1-\eta\eta'(\sigma)\right)\sqrt{\det V_\sigma}\\
        z_0 &:= \eta \sqrt{\det V_\rho}+ (1-\eta)\sqrt{\det V_\xi}\\
        z_1 &:= 2(2f(z_0)-1)\sqrt{\det V_\rho}-z_0.
    \end{align}
\end{rst}

\begin{proof}
        Let $\rho$ and $\sigma$ be  thermal states, i.e., $V_{\rho} = (2\overline{n}_{\rho}+1)I, \overline{\mathbf{r}}_\rho=0$ and $V_{\sigma} = (2\overline{n}_{\sigma}+1)I, \overline{\mathbf{r}}_\sigma=0$, respectively. Recall that the lossy channel under study does not have a displacement effect, i.e., $\mathbf{d}_\mathcal{N}=0$. Together with the fact that $\overline{\mathbf{r}}_\sigma=0$,
        we have $\mathbf{d}_\mathcal{P}=0$. Collectively they lead to 
    \begin{align*}
        F(\rho,\mathcal{N}(\rho)) = \frac{1}{\Gamma(V_\rho,V_{\mathcal{N}(\rho)})} \qand F(\rho,\mathcal{P}\circ\mathcal{N}(\rho)) = \frac{1}{\Gamma(V_\rho,V_{\mathcal{P}\circ\mathcal{N}(\rho)})}.
    \end{align*}
    
    Here, we will show $\Gamma(V_\rho,V_{\mathcal{N}(\rho)}) \ge \Gamma(V_\rho,V_{\mathcal{P}\circ\mathcal{N}(\rho)})$ to prove $F(\rho, \mathcal{N}(\rho)) \le F(\rho, \mathcal{P}_{\mathcal{N},\sigma}\circ \mathcal{N}(\rho))$.
    Using Eq.~\eqref{eq:Delta} and Eq.~\eqref{eq:Lambda}, we have
    \begin{align}
        &\Delta(V_\rho,V_{\mathcal{N}(\rho)}) = \frac{1}{4} \det(V_\rho+V_{\mathcal{N}(\rho)}), \quad  \Lambda(V_\rho,V_{\mathcal{N}(\rho)}) = \frac{1}{4}(\det V_\rho-1)(\det V_{\mathcal{N}(\rho)}-1),\\
        &\Delta(V_\rho,V_{\mathcal{P}\circ\mathcal{N}(\rho)}) = \frac{1}{4} \det(V_\rho+V_{\mathcal{P}\circ\mathcal{N}(\rho)}), \qand  \Lambda(V_\rho,V_{\mathcal{P}\circ\mathcal{N}(\rho)}) = \frac{1}{4}(\det V_\rho-1)(\det V_{\mathcal{P}\circ\mathcal{N}(\rho)}-1).
    \end{align}

        We first note that $V_{\mathcal{N}(\rho)}=\eta V_\rho+(1-\eta)V_\xi$, and thus 
    \begin{align}
        V_{\mathcal{P}\circ\mathcal{N}(\rho)} &=\eta'V_{\mathcal{N}(\rho)} + Y_\mathcal{P}\\
        &=\eta'V_{\mathcal{N}(\rho)} + V_\sigma-\eta' V_{\mathcal{N}(\sigma)}\\
        &= \eta\eta'V_\rho + \left(1-\eta\eta'\right)V_\sigma, \label{eq:VPNrho}
    \end{align}
    where we used Eq.~\eqref{eq:Petz_eta'} in the first equation and the fact that $Y_\mathcal{P}=V_{\sigma}-X_\mathcal{P}V_{\mathcal{N}(\sigma)}X_\mathcal{P}^T=V_{\sigma}-\eta'V_{\mathcal{N}(\sigma)}$ in the second equation. Denoting 
    \begin{align}
        x:=(\det V_\sigma)^\frac{1}{2} = 2 \overline{n}_\sigma+1,    
    \end{align}
    we will identify $\Delta$'s and $\Lambda$'s as functions of $x$. We further denote  
    \begin{align}
        a:=(\det V_\rho)^\frac{1}{2} = 2 \overline{n}_\rho+1 \qand b:=(\det V_\xi)^\frac{1}{2} = 2 \overline{n}_\xi+1.
    \end{align}
    We get 
    \begin{align}
        \det V_{\mathcal{N}(\sigma)} &= \det\left( \eta V_\sigma + (1-\eta)V_\xi\right)\\
        &= (\eta x+ (1-\eta)b)^2.\\
        \det V_{\mathcal{N}(\rho)} &= \det\left( \eta V_\rho + (1-\eta)V_\xi\right)\\
        &= (\eta a+ (1-\eta)b)^2.
    \end{align}
    Moreover, $\eta'$ is also a function of $x$, i.e., 
    \begin{align}
        \eta'=\eta'(x):=\eta\frac{x^2-1}{(\eta x+ (1-\eta)b)^2-1}.
    \end{align}
    $\det V_{\mathcal{P}\circ\mathcal{N}(\rho)}$ and $\det\left(V_\rho + V_{\mathcal{P}\circ\mathcal{N}(\rho)}\right)$ are the key ingredients to compute $\Gamma(V_\rho,V_{\mathcal{P}\circ\mathcal{N}(\rho)})$, and they can be recast as
    \begin{align*}
        &\det V_{\mathcal{P}\circ\mathcal{N}(\rho)} \nonumber = \left(\eta\eta'(x)a + \left(1-\eta\eta'(x)\right)x \right)^2=:g(x)^2 \qand \\
        &\det\left(V_\rho + V_{\mathcal{P}\circ\mathcal{N}(\rho)}\right) = \left((1+\eta\eta'(x))a + \left(1-\eta\eta'(x)\right)x \right)^2 =(a+g(x))^2,
    \end{align*}
    where we introduced a function $g(x)$ for notational simplicity, defined as 
    \begin{align}
        g(x):= \eta\eta'(x)a + \left(1-\eta\eta'(x)\right)x.
    \end{align}
    
    In turn, we can identify $\Delta$'s and $\Lambda$'s as functions of $x$ as follows:
        \begin{align}
        \Delta(V_\rho,V_{\mathcal{N}(\rho)}) &= \frac{1}{4} (a+\eta a+ (1-\eta)b)^2, \\ \Lambda(V_\rho,V_{\mathcal{N}(\rho)}) &= \frac{1}{4}(a^2-1)\left((\eta a+ (1-\eta)b)^2-1\right),\\
        \Delta(V_\rho,V_{\mathcal{P}\circ\mathcal{N}(\rho)}) &= \frac{1}{4} \left(a+g(x)\right)^2, \,\text{and} \\ \Lambda(V_\rho,V_{\mathcal{P}\circ\mathcal{N}(\rho)}) &= \frac{1}{4}(a^2-1)\left(g(x)^2-1\right).
    \end{align}

    Defining a function 
    \begin{align}
        f(z):=\frac{1}{2}\left( az+1 - (a^2-1)^\frac{1}{2}(z^2-1)^\frac{1}{2} \right),
    \end{align}
    we notice that $\Gamma(V_\rho,V_{\mathcal{N}(\rho)})$ and $\Gamma(V_\rho,V_{\mathcal{P}\circ\mathcal{N}(\rho)})$ can be expressed as
    \begin{align}
        \Gamma(V_\rho,V_{\mathcal{N}(\rho)}) &=f(\eta a+ (1-\eta)b),\\
        \Gamma(V_\rho,V_{\mathcal{P}\circ\mathcal{N}(\rho)}) &= f(g(x)).
    \end{align}
    Hence, we arrive at 
    \begin{align}
        &F(\rho, \mathcal{N}(\rho)) \le F(\rho, \mathcal{P}_{\mathcal{N},\sigma}\circ \mathcal{N}(\rho)) \nonumber\\
        &\Leftrightarrow \Gamma(V_\rho,V_{\mathcal{N}(\rho)}) \ge \Gamma(V_\rho,V_{\mathcal{P}\circ\mathcal{N}(\rho)}) \\
        &\Leftrightarrow f(\eta a+ (1-\eta)b)\ge f(g(x)).
        \label{eq:con_r0}
    \end{align}

    Denote $z_0 = \eta a+ (1-\eta)b$ for further simplification. Notice that function $f(z)$ is continuous and “parabola-like”, the condition \cref{eq:con_r0} corresponds to the function lying below a horizontal line. Therefore, $g(x)$ should lie between two solutions of the equation $f(z) = f(z_0)$. That is, to solve 
    \begin{equation}
        \begin{split}
            &\frac{1}{2}\left( az+1 - (a^2-1)^\frac{1}{2}(z^2-1)^\frac{1}{2} \right) = f(z_0)\\
            & \Leftrightarrow ((a^2-1)^{1/2}(z^2-1)^{1/2})^2 = (az+1-2f(z_0))^2\\
            & \Leftrightarrow z^2 -2(2f(z_0)-1)az +(2f(z_0)-1)^2+a^2-1=0. 
        \end{split}
    \end{equation}
    By Vieta's theorem, the other solution satisfies relation $z_0 + z_1 = 2(2f(z_0)-1)a$. 
    At the same time, function $f(z)$ is defined with $z \geq 1$. Combining these all together, $F(\rho, \mathcal{R}^{(0)}\circ\mathcal{N}(\rho))\equiv F(\rho, \mathcal{N}(\rho)) \le F(\rho, \mathcal{P}_{\mathcal{N},\sigma}\circ \mathcal{N}(\rho))$ leads to $\max(1, \min (z_0,z_1)) \le g(\sigma) \leq \max (z_0, z_1)$, where $z_0 = \eta \sqrt{\det V_\rho} + (1-\eta) \sqrt{\det V_\xi}$, $z_1 = 2(2f(z_0)-1)\sqrt{\det V_\rho}-z_0$. 
\end{proof}

\subsection{The recovery protocol $\mathcal{R}^{(1)}_{\sigma}$}\label{app:fidelity1}

\begin{rst}
    Let $\rho$ and $\sigma$ be thermal states. Then, we have 
    \begin{align}
        F(\rho, \mathcal{R}^{(1)}_\sigma\circ\mathcal{N}(\rho))= F(\rho, \sigma) \le F(\rho, \mathcal{P}_{\mathcal{N},\sigma}\circ \mathcal{N}(\rho)).
    \end{align}
\end{rst}
\begin{proof}
    Let $\rho$ and $\sigma$ be thermal states, i.e., $V_{\rho} = (2\overline{n}_{\rho}+1)I, \overline{\mathbf{r}}_\rho=0$ and $V_{\sigma} = (2\overline{n}_{\sigma}+1)I, \overline{\mathbf{r}}_\sigma=0$, respectively. Similar to the previous proof, the fidelity can be simplified to 
    \begin{align*}
        F(\rho,\sigma) = \frac{1}{\Gamma(V_\rho,V_\sigma)} \qand F(\rho,\mathcal{P}\circ\mathcal{N}(\rho)) = \frac{1}{\Gamma(V_\rho,V_{\mathcal{P}\circ\mathcal{N}(\rho)})},
    \end{align*}
    where $\Gamma$ is defined as in Eq.~\eqref{eq:Gamma}.
    
    We will therefore show $\Gamma(V_\rho,V_\sigma) \ge \Gamma(V_\rho,V_{\mathcal{P}\circ\mathcal{N}(\rho)})$ to prove $F(\rho, \sigma) \le F(\rho, \mathcal{P}_{\mathcal{N},\sigma}\circ \mathcal{N}(\rho))$. Here we will use the similar techniques that were used in the previous proof. Using the same notations as in the previous proof, i.e.,  
    \begin{align}
        &x:=(\det V_\sigma)^\frac{1}{2}, \quad a:=(\det V_\rho)^\frac{1}{2} \quad b:=(\det V_\xi)^\frac{1}{2} ,\\
        &\eta'(x):=\eta\frac{x^2-1}{(\eta x+ (1-\eta)b)^2-1} \quad g(x):= \eta\eta'(x)a + \left(1-\eta\eta'(x)\right)x,
    \end{align}
    we then get 
    \begin{align}
        \Gamma(V_\rho,V_\sigma) &=\frac{1}{2}\left( ax+1 - (a^2-1)^\frac{1}{2}(x^2-1)^\frac{1}{2} \right)=:f(x),\\
        \Gamma(V_\rho,V_{\mathcal{P}\circ\mathcal{N}(\rho)}) &=  \frac{1}{2}\left( ag(x)+1 - (a^2-1)^\frac{1}{2}(g(x)^2-1)^\frac{1}{2} \right)=:f(g(x)).
    \end{align}
    
    Observe now that the derivative of $f(x)$ is always positive for $x\ge a$:
    \begin{align}
        \frac{\dd{f}}{\dd{x}} = \frac{\sqrt{a^2-1}}{2}\left( \frac{a}{\sqrt{a^2-1}} - \frac{x}{\sqrt{x^2-1}} \right)\ge0,
    \end{align}
    because $y=\frac{x}{\sqrt{x^2-1}}$ is monotonically decreasing for $x\ge a>0$. That is, $f(x)$ is monotonically increasing for $x\ge a$. Hence, if $x\ge a$, we have 
    \begin{align}
        &F(\rho, \sigma) \le F(\rho, \mathcal{P}_{\mathcal{N},\sigma}\circ \mathcal{N}(\rho)) \nonumber\\
        &\Leftrightarrow \Gamma(V_\rho,V_\sigma) \ge \Gamma(V_\rho,V_{\mathcal{P}\circ\mathcal{N}(\rho)}) \\
        &\Leftrightarrow f(x)\ge f(g(x)) \\
        &\Leftrightarrow g(x)-x \le 0.
    \end{align}

    Meanwhile, notice $g(x)-x\le 0$ for $x\ge a$ because $g(x)-x = \eta\eta'(x)(a-x)$ and $\eta\ge 0, \eta'(x) \ge 0$. 
    
    Conversely, if $x<a$, $f(x)$ is monotonically decreasing, thus we need to show $g(x)-x\ge 0$ to prove $F(\rho, \sigma) \le F(\rho, \mathcal{P}_{\mathcal{N},\sigma}\circ \mathcal{N}(\rho))$. We complete the proof since $g(x)-x=\eta\eta'(x)(a-x)\ge 0$ for $x<a$.
\end{proof}

\section{Simplification of two-mode Petz recovery map}
\label{two mode}

In this section, we consider the two mode symmetric lossy channel expressed as two beam splitters with identical transmissivity $0 \leq \eta \leq 1$, and two identical thermal environment state with mean photon number $\overline{n}_{\xi}$. The forward channel can be expressed by
\begin{align}
    X_{\mathcal{N}_{tot}} = \sqrt{\eta}I_4, Y_{\mathcal{N}_{tot}} = (1-\eta) (V_\xi \oplus V_\xi), \mathbf{d}_{\mathcal{N}_{tot}} = 0,
\end{align}
where $I_4$ is the $4 \times 4$ identity matrix, and $V_\xi = (2\overline{n}_\xi+1)I_2$ is the covariance matrix of thermal environment state. 

Now we assume that the reference state of the Petz recovery map is a symmetric squeezed thermal state, whose covariance matrix is given by 
\begin{align}
    V_\sigma = (2\overline{n}_\sigma+1) \left [ 
    \begin{array}{cc}
        \cosh 2r I_2 & -\sinh 2r S_0\\
        -\sinh 2r S_0 & \cosh 2r I_2
    \end{array}
    \right ], \text{ with } S_0 = \left [
    \begin{array}{cc}
        1 & 0 \\
        0 & -1
    \end{array}
    \right ]. 
\end{align}
$\overline{n}_\sigma$ is the mean photon number of each thermal state, and $r$ is the squeezing parameter. 

Following \cref{eq:petz_XY}, the transformation matrix of the Petz recovcovery map becomes
\begin{align}
    \begin{aligned}
        X_\mathcal{P} &= (I_4 + (V_\sigma \Omega )^{-2})^{1/2} V_\sigma X_{\mathcal{N}_{tot}}^T (I_4 + (\Omega V_{\mathcal{N}_{tot}(\sigma)})^{-2})^{-1/2} V_{\mathcal{N}_{tot}(\sigma)}^{-1}\\
        &= (I_4 - (\det V_\sigma) ^{-1/2} I_4)^{1/2} V_\sigma \sqrt{\eta} I_4 (I_4- (\det V_{\mathcal{N}_{tot}(\sigma)})^{-1/2}I_4)^{-1/2} V^{-1}_{\mathcal{N}_{tot}(\sigma)}\\
        &= \sqrt{\eta} \sqrt{\frac{1-(\det V_\sigma)^{-1/2}}{1-(\det V_{\mathcal{N}_{tot}(\sigma)})^{-1/2}}} V_\sigma V^{-1}_{\mathcal{N}_{tot}(\sigma)}. 
    \end{aligned}
\end{align}
Here we use the simplifications:
\begin{enumerate}
    \item For a $4 \times 4$ matrix that is proportional to squeezing matrix $S(r)$, i.e., $M \propto S(r)$, $M\Omega M \Omega = (\det M)^{1/2}I_4$.
    \item For a $4 \times 4$ matrix that has shape 
    \begin{align}
        M = \left [
            \begin{array}{cc}
               aI_2  & c S_0 \\
                c S_0 & bI_2
            \end{array}
            \right], 
    \end{align}
    $\Omega M \Omega M =  (\det M)^{1/2}I_4$. 
\end{enumerate}
Now since $V_\sigma V^{-1}_{\mathcal{N}_{tot}(\sigma)}$ has shape 
$\left [
\begin{array}{cc}
    AI & BS_0 \\
    BS_0 & AI
\end{array}
\right ]$, it is possible to decompose the transformation matrix of the Petz recovery map to another squeezing matrix $S(r')$, with a coefficient to be another set of beam splitters or phase-insensitive amplifiers. 

Let us decompose the transformation matrix as 
\begin{align}
    \label{eq:decomposition}
    X_\mathcal{P} \equiv X(\eta') S(r'),
\end{align}
where $X(\eta') \equiv \sqrt{\eta'} I_4$ is the transformation matrix of beam splitter $(0\leq \eta' \leq 1)$ or phase-insensitive amplifier $\eta' >1$. We further express $S(r') \equiv \sqrt{\chi} V_\sigma V^{-1}_{\mathcal{N}_{tot}(\sigma)}$, where $\chi$ is the coefficient such that $\det S(r') = 1$. 

We first work out the expression of $\chi$. By having $\det S(r') = 1$, we get 
\begin{align}
    \begin{aligned}
        \det S(r') &= \chi^2 \det (V_\sigma V^{-1}_{\mathcal{N}_{tot}(\sigma)} )\\
        &= \chi^2 \det V_\sigma \det V^{-1}_{\mathcal{N}_{tot}(\sigma)}\\
        &= 1. 
    \end{aligned}
\end{align}
So we get 
\begin{align}
    \chi = \sqrt{\frac{\det V_{\mathcal{N}_{tot}(\sigma)}}{\det V_\sigma}}. 
\end{align}

Back to \cref{eq:decomposition}, we will have $X_\mathcal{N} = X(\eta')S(r') = \sqrt{\eta'} \sqrt{\chi} V_\sigma V^{-1}_{\mathcal{N}_{tot}(\sigma)} = \sqrt{\eta \frac{1-(\det V_\sigma)^{-1/2}}{1-(\det V_{\mathcal{N}(\sigma)})^{-1/2}}} V_\sigma V_{\mathcal{N}(\sigma)}^{-1}$, from which we get the expression of $\eta'$:
\begin{align}
    \begin{aligned}
        \eta' &= \eta \frac{1-(\det V_\sigma)^{-1/2}}{1-(\det V_{\mathcal{N}(\sigma)})^{-1/2}} \frac{1}{\chi}\\
        &= \eta \frac{1-(\det V_\sigma)^{-1/2}}{1-(\det V_{\mathcal{N}(\sigma)})^{-1/2}} \frac{(\det V_\sigma)^{1/2}}{(\det V_{\mathcal{N}(\sigma)})^{1/2}}. 
    \end{aligned}
\end{align}

Finally we will determine the squeezing parameter $r'$. By having $V_\sigma V^{-1}_{\mathcal{N}_{tot}(\sigma)} \propto S(r')$ with $V_\sigma V_{\mathcal{N}_{tot}(\sigma)}$  having shape 
$\left [
\begin{array}{cc}
    AI & BS_0 \\
    BS_0 & AI
\end{array}
\right ]$, we will have 
\begin{align}
    \frac{B}{A} &= -\tanh r', 
\end{align}
where $A = \frac{\Tilde{n}_\sigma (\eta \Tilde{n}_\sigma  + (1-\eta)\Tilde{n}_\xi \cosh(2r_\sigma))}{(1-\eta)^2 \Tilde{n}_\xi^2  + \eta^2 \Tilde{n}_\sigma^2  + 2(1-\eta) \eta  \Tilde{n}_\sigma \Tilde{n}_\xi  \cosh(2r_\sigma)}$, $B =- \frac{(1-\eta) \Tilde{n}_\sigma \Tilde{n}_\xi  \sinh(2r_\sigma)}{(1-\eta)^2 \Tilde{n}_\xi^2  + \eta^2 \Tilde{n}_\sigma^2  +2 (1-\eta)\eta   \Tilde{n}_\sigma \Tilde{n}_\xi \cosh(2r_\sigma)}$. Here $\Tilde{n}_{\sigma,\xi} = 2\overline{n}_{\sigma,\xi}+1$. From here we get 
\begin{align}
    r' = \arctanh \left (\frac{(1-\eta)(2\overline{n}_\xi+1)}{\eta(2\overline{n}_\sigma+1)+(1-\eta)(2\overline{n}_\xi+1)} \tanh (2r_\sigma)
    \right ). 
\end{align}

From above, we get the decomposition of the Petz recovery map.

\end{document}